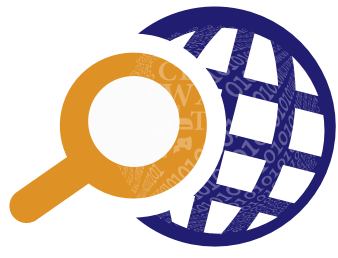

# جامعة الأمير سلطان

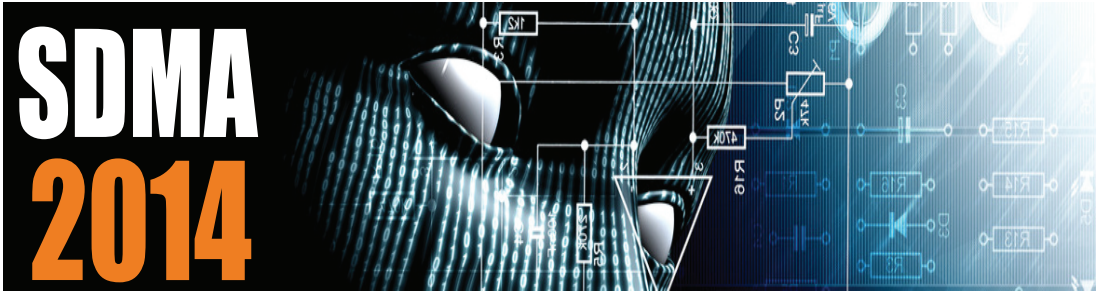

# The Third SYMPOSIUM On
# DATA MINING APPLICATIONS

Prince Megren Data Mining Center (MEGDAM), Prince Sultan University (PSU), Riyadh, Saudi Arabia

Thursday 8th May, 2014

# TECHNICAL PROCEEDINGS

# SDMA 2014

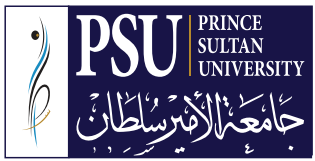
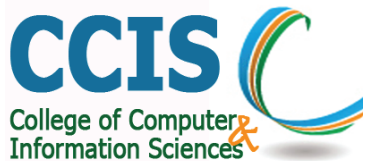
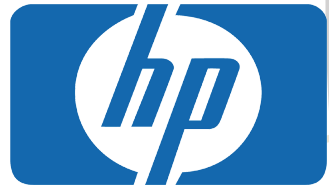
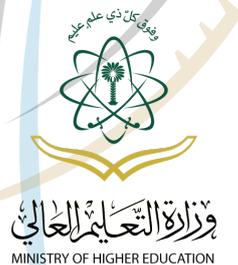

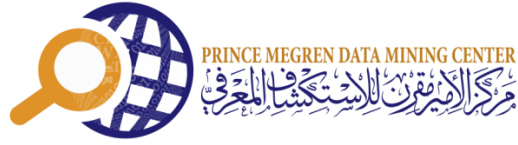
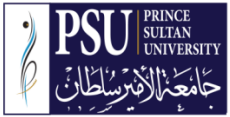 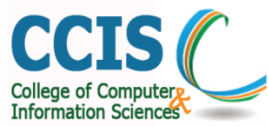 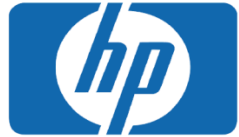 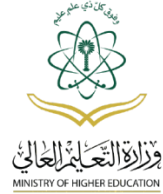

Published By

Prince Megrin Data Mining Center, PSU

SDMA 2014

Proceedings of Symposium on Data Mining Applications 2014

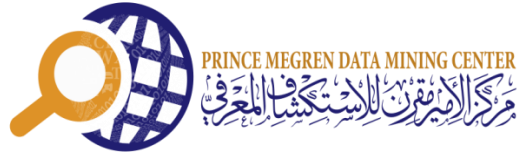
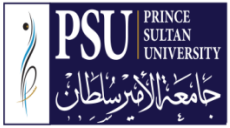 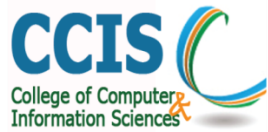 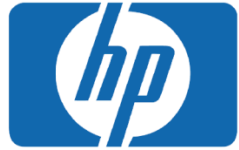 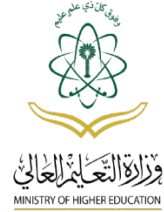

# Third Symposium on Data Mining Applications 2014
# Thursday 8th May 2014
# Prince Sultan University, Riyadh, K.S.A

**General Chair**

Dr. Abdul Aziz Sehibani, PSU, KSA

**Steering Committee**

Dr. Basit Qureshi, PSU, KSA
Dr. Mohamed Tounsi, PSU, KSA

**Program Committee**

Dr. Mohammad El Affendi , PSU, KSA
Dr. Ahmed Sameh, PSU, KSA
Dr. Izzat AlSmadi, PSU, KSA
Dr. Ajantha Dhanyake, PSU, KSA
Dr. Nour, PSU, KSA
Dr. Soufianne Nouredin, PSU, KSA
Dr. Manal Farrag, PSU, KSA
Dr.Liyakathunisa, PSU, KSA
Dr. Tanzila Saba, PSU, KSA

Dr. Afraz Zahra Syed, PSU, KSA
Ms. Fatima Khan, PSU, KSA
Dr. Ibrahim Abu Nadi, PSU, KSA
Dr. Fakhry Khellah, PSU, KSA
Dr Anis Koubaa, PSU, KSA
Dr Anis Zarrad, PSU, KSA
Dr. Romana Aziz, PSU, KSA
Dr Suad AlRamouni, PSU, KSA

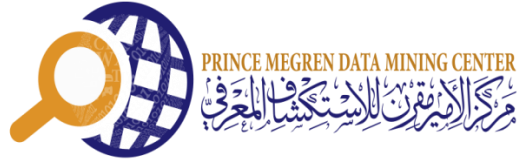
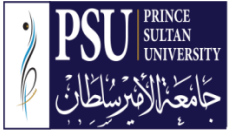 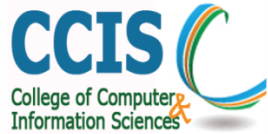 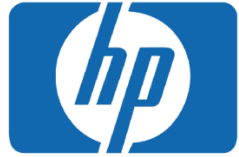 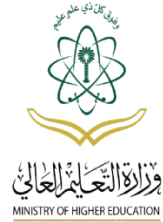

# Third Symposium on Data Mining Applications 2014
# Thursday 8<sup>th</sup> May 2014
# Prince Sultan University, Riyadh, K.S.A

## Foreword

The Symposium on Data Mining and Applications (SDMA 2014) is aimed to gather researchers and application developers from a wide range of data mining related areas such as statistics, computational intelligence, pattern recognition, databases, Big Data Mining and visualization. SDMA is organized by MEGDAM to advance the state of the art in data mining research field and its various real world applications. The symposium will provide opportunities for technical collaboration among data mining and machine learning researchers around the Saudi Arabia, GCC countries and Middle-East region. Acceptance will be based primarily on originality, significance and quality of contribution.

Topics for this symposium include, but are not limited to:

**Core Data Mining Topics**

| | |
|---|---|
| Parallel and distributed data mining algorithms | Pre-processing techniques |
| Data streams mining | Visualization |
| Graph mining | Security and information hiding in data mining |
| Spatial data mining | |
| Text video, multimedia data mining | |
| Web mining | |
| Big Data mining | |

**Data Mining Applications**

| | |
|---|---|
| Databases | Classification |
| Healthcare | Clustering |
| Financial | Networks |
| Security | Educational data mining |
| Forecasting | Bioinformatics |

SDMA 2014 also includs invited talks that are as follows

- **Mining Patterns from Complex Datasets via Sampling**
  Mohammad Zaki - QCRI

- **Using 10,000 Processors to Extract Frequent Patterns from Very Long Sequences**
  Panos Kalinis - KAUST

- **Arabic Processing and Text Mining**
  Husni Al Muhtaseb - KFUPM

- **A Computer Aided Diagnosis System for Lung Cancer based on Statistical and Machine Learning Techniques**
  Samir Belahouri – AlFaisal University

- **Opinion Mining and Sentimental Analysis for Arabic**
  Izzat M. AlSamadi - PSU

- **Statistical Parsing**
  Mohammad El Affendi - PSU

- **Maximize business outcomes with HP Data Mining**
  Hamid Wassifi - HP

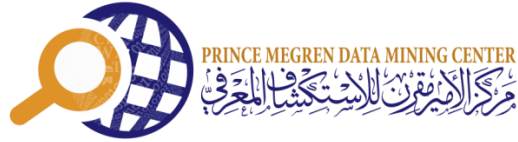
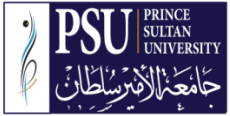 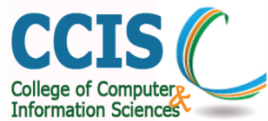 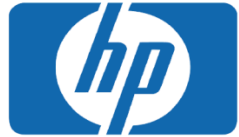 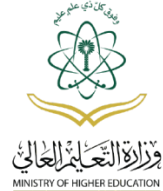

# Third Symposium on Data Mining Applications 2014
# Thursday 8[th] May 2014
# Prince Sultan University, Riyadh, K.S.A

## Invited Speakers Bio and Talk Abstract

| Mining Patterns from Complex Datasets via Sampling | Mohammad Zaki |
|---|---|

**Abstract:** Data, data, everywhere, but not any drop of insight! Many real world problems require the discovery of complex patterns from very large datasets. However, taming the complexity and scalability challenges is essential for knowledge discovery. In this talk, I will highlight some of our recent work on mining complex patterns such as attributed subgraphs, boolean expressions, and so on, based on sampling strategies.

**Speaker Bio:** Mohammed J. Zaki is a Professor of Computer Science at Rensselaer Polytechnic Institute (RPI), Troy, NY. He is also a Principal Scientist at the Qatar Computing Research Institute (QCRI) in Doha, Qatar. He received his Ph.D. degree in computer science from the University of Rochester in 1998. His research interests focus on developing novel data mining techniques, especially for social networks and bioinformatics. He has over 200 publications in data mining and bioinformatics. He is currently Area Editor for Statistical Analysis and Data Mining, and an Associate Editor for Data Mining and Knowledge Discovery, ACM Transactions on Knowledge Discovery from Data, Knowledge and Information Systems, Social Networks and Mining, and International Journal of Knowledge Discovery in Bioinformatics. He was the program co-chair for SDM'08, SIGKDD'09, PAKDD'10, BIBM'11, CIKM'12, and ICDM'12, and he is the founding co-chair for the BIOKDD workshop. He received several research awards, such as the NSF CAREER Award, DOE Career Award, HP Innovation Research Award, and Google Faculty Research Award. He is a senior member of the IEEE, and an ACM Distinguished Scientist.

## Using 10,000 Processors to Extract Frequent Patterns from Very Long Sequences — Panos Kalinis

**Abstract:** Modern applications, including bioinformatics, time series, and web log analysis, require the extraction of frequent patterns, called motifs, from one very long sequence. Existing approaches are either heuristics that are error-prone; or exact combinatorial methods that are extremely slow; therefore applicable only to very small sequences, in the order of megabytes.

This talk introduces ACME, a combinatorial approach for motif extraction that scales to gigabyte long sequences. ACME is a versatile parallel system that can be deployed on desktop multi-core systems, or on thousands of CPUs in the cloud. However, merely using more compute nodes does not guarantee efficiency, because of the related overheads. To this end, ACME introduces an automatic tuning mechanism that suggests the appropriate number of CPUs to utilize, in order to meet the user constraints in terms of run time, while minimizing the financial cost of cloud resources. Our experiments show that, compared to the state of the art, ACME: (i) supports 3 orders of magnitude longer sequences (e.g., DNA for the entire human genome); (ii) handles large alphabets (e.g., English alphabet for Wikipedia); (iii) scales out to 16,384 CPUs on a supercomputer; and (iv) is the only method to support elastic deployment on the cloud.

**Speaker Bio**: Panos Kalnis is associate professor of Computer Science in the King Abdullah University of Science and Technology (KAUST). He is leading the InfoCloud group that focuses on Big Data. In 2009 he was visiting assistant professor in Stanford University. Before that, he was assistant professor in the National University of Singapore (NUS). In the past he was involved in the designing and testing of VLSI chips in the Computer Technology Institute, Greece. He also worked in several companies on database designing, e-commerce projects and web applications. He is an associate editor for the IEEE Transactions on Knowledge and Data Engineering and serves on the editorial board of The VLDB Journal. He received his Diploma in Computer Engineering from the Computer Engineering and Informatics Dept., University of Patras, Greece in 1998 and his PhD from the Computer Science Dept., Hong Kong University of Science and Technology (HKUST) in 2002. His research interests include Big Data, Analytics, Cloud Computing, Distributed Systems, Large Graphs and Sequences.

## Arabic Processing and Text Mining — Husni Al Muhtaseb

**Abstract:** Arabic is the first language for more than 400 million people in the world. It is also used by more than triple the previous number of Muslims all over the world as a second language, for it is the language in which the Holy Qur'an was revealed. That is, more than 1.5 billion people are using Arabic. Processing Arabic by computers is important for proper use in a wide range of technologies and research areas in Human Language Technology. These areas include Natural Language Processing (NLP), Speech Recognition, Machine Translation, Text Generation and Text Mining. Until now, Arabic language is not well-served by electronic resources to make text mining facilitated. More research work is needed to help lay the foundation for text mining in Arabic. In this seminar, we focus on needed resources to process Natural Arabic for text mining.

**Speaker Bio** : Husni Al-Muhtaseb Obtained a PhD degree from the Department of Electronic Imaging and Media Communications (EIMC) of the School of Informatics in the University of Bradford, UK in 2010. He received his M.S. degree in computer science and engineering from King Fahd University of Petroleum and Minerals (KFUPM), Dhahran, Saudi Arabia, in 1988 and the B.E. degree in electrical engineering, computer option, from Yarmouk University, Irbid, Jordan in 1984. He is currently an Assistant Professor of Information and Computer Science at KFUPM. His research interests include software development, Arabic Computing, computer Arabization, e-learning & online tutoring and natural Arabic understanding.  He is a member of Association of Jordanian Engineers, Electrical Engineering Division and Saudi Computer Society. He developed the first course in Arabization of Computers in the world. The course is now being taught in 10s of Universities and colleges. He has participated in several industrial projects with different institutes/ organizations including, KACST, STC, MOHE and Aramco. He also worked as a consultant for different entities including KFUPM schools and Ministry of education.

## A Computer Aided Diagnosis System for Lung Cancer based on Statistical and Machine Learning Techniques — Samir Belahouri

**Abstract :** Lung Cancer is believed to be among the primary factors for death across the world. Within this paper, statistical and machine learning techniques are employed to build a computer aided diagnosis system for the purpose of classifying lung cancer. The system includes preprocessing phase, feature extraction phase, feature selection phase and classification phase. For feature extraction, wavelet transform is used and for feature selection, two-step statistical techniques are applied. Clustering-K-nearest-neighbor classifier is employed for classification. The Japanese Society of Radiological Technology's standard dataset of lung cancer has been utilized to evaluate the system. The dataset has 154 nodule regions (abnormal) - where 100 are malignant and 54 are benign - and 92 non-nodule regions

(normal). An Accuracy of 99.15% and 98.70 % for classification have been achieved for normal versus abnormal and benign versus malignant respectively, this substantiate the capabilities of the approach

**Speaker Bio :** Samir Brahim Belhaouari is the Head of Mathematics Department at College of Science, ALFAISAL University. He earned his Ph.D. in Mathmatical Science from Federal Polytechnic School of Lausanne in Switzerland, MS., in Networks and Telecommunications from National Polytechnic Institutes in France. His research expertise are a combination of Applied Mathematics, Patter Recognition & Prediction techniques, Image & Signal Processing (Biomedical, Bioinformatics, Forecasting, Communication…), Artificial Intelligence, Statistics, and Random Walks . He has achieved many honors and recognitions throughout his career. He has also a member in many editorial boards for international journals and a reviewer. He has published more than 145 (combination of books, journals, conferences … etc.)

## Statistical Parsing                  Muhammad El-Affendi

**Abstract :** Parsing is an important component of most serious Natural Language Processing (NLP) applications. Like other components in NLP systems, workers in this field are benefiting a lo from the large corpora that has been built over the past few years. In this session we reflect briefly on the developments in the area of statistical parsing.

**Speaker Bio :** Dr. Mohammad El Affendi is Professor and Dean College of Computer and Information Sciences at Prince Sultan University (PSU). He received his M.Sc. and Ph.D. in computer Science from the University of Bradford, U.K. He Founded Prince Salman Research Center at PSU, and prepared its strategic plan and directed it for 3 years. He participated in many of consultancy and strategic planning activities. He has more than 30 years of experience in Administrative and Directing in Computer Science related fields. He is also very active in research with more than 35 publications in such diverse areas: Distributed applications & systems, Speech and Natural Language Processing, Web & Cluster processing and software engineering. He is currently leading a research project on Arabic NLP at Prince Megrin Data Mining Center.

## Opinion Mining and Sentimental Analysis for Arabic      Izzat Al Samadi

**Abstract :** The Internet continuously evolves to include lots of new features for serving and communicating with people and users. Recently the web or most websites include the ability for normal users to contribute to the Internet and web content not only based on building personal websites or pages but to include their comments, posts, or feedbacks on posts, articles, products, etc.

This can largely be categorized into:

1.      E-commerce websites (e.g. Amazon) to allow users to post their feedback on the books or products that they bought and rate the product and the service.

2.      News websites to allow users t post their comments related to the specific article.

3.      Social networks where individual can create their own posts or responds to their friends ot others' posts.

The presentation will cover issues related to sentimental analysis in particular in Arabic language and its relation with: Information Retrieval, Natural Language Processing, and data or text mining.

**Speaker Bio :** Dr. Izzat M. AlSamadi is an Associate Professor at the Department of Computer Information Systems, and Member of WINS (Wireless Information Networks and Security) Research Group at PSU. He obtained his MS and Ph.D. in Software Engineering from North Dakota State University (NDSU), USA. He obtained his BS degree in Computer Information Systems from University of Phoenix, USA and his BS degree in Telecommunication Engineering from Mutah University, Jordan. He has authored or co-authored more than 100 publications in different international journals and conferences. He is actively working in Arabic text mining and NLP.

## Maximize business outcomes with HP Data Mining — Hamid Wassifi

**Abstract :** Understanding Information meaning is the key to solving information challenges and the basis of any business decision at any level and for any purpose. Volume of human friendly information produced, captured, exchanged is growing exponentially. For this reason organization needs the right tools to Govern, Protect, Understand, Analyze, Share, Present this information and to extract the maximum value and benefits out of it. HP through a Meaning Based Computing approach help organization to deliver the right information, at the right time, to the right people, to support enterprise outcomes.

**Speaker Bio**: Hamid Wassifi has been working in Software industry for 20 years across Asia, Europe, Middle East and Africa to design, implement and provide consulting on Software solution adapted to client expectations. He has been in his present role at HP Autonomy for 2 years focusing on Information Management and Analytics solution

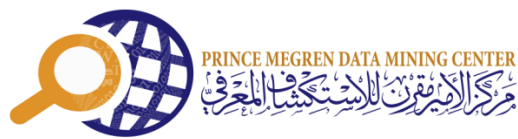
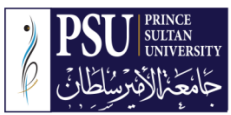
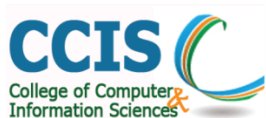
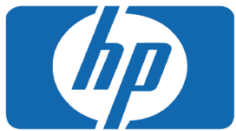
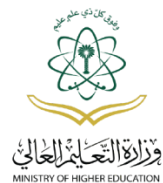

# Third Symposium on Data Mining Applications 2014

Thursday 8th May 2014

Prince Sultan University, Riyadh, K.S.A

## List of Authors

| Name | Paper No |
| --- | --- |
| Adel Aloraini | P02 |
| Adel Aloraini | P10 |
| Amani Al-Ghanayem | P01 |
| Arwa Al-Rofiyee | P03 |
| Basit Qureshi | P08 , P12 |
| Bayan Almkhelfy | P10 |
| Bedoor Al-Saeed | P11 |
| Fahad Almudarra | P08 |
| Faten Alrusayni | P02 |
| Izzat Alsmadi | P04 |
| Maram Al-Nowiser | P03 |
| Mohammad Zarour | P04 |
| Mohammed Abdullah AL-Hagery | P03, P11 |
| Muhammad Badruddin Khan | P06 , P07 , P09 |
| Nasebih Al-Mufadi | P03 |
| Nouf Almutairi | P05 |
| Nourh Abdulaziz Rasheed | P06 |
| Rehab Nasser Al-Wehaibi | P07 |
| Riyad Alshammari | P05 |
| Salha al Osaimi | P09 |
| Samar Al-Qarzaie | P11 |
| Sara Al-Odhaibi | P11 |
| Waleed Rashaideh | P01 |
| Yasir Javed | P12 |

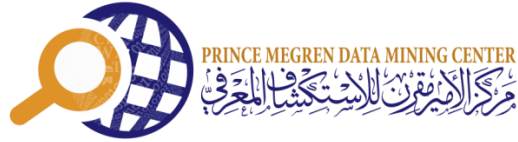

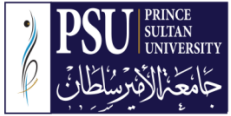 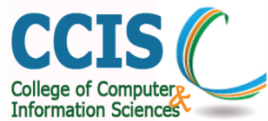 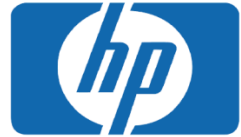 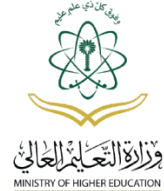

# SDMA 2014

Third Symposium on Data Mining Applications 2014

Thursday 8th May 2014

Prince Sultan University, Riyadh, K.S.A

# Technical Procedings

# Identifying Word(s) Tendencies in Classification of Arabic News Using Text Mining Techniques

Amani, Al-Ghanayem*, Dr.Waleed, Rashaideh †
* College of Computer and Information Science, Al-Imam Muhammad Ibn Saud Islamic University, Saudi Arabia
†College of Computer and Information Science, Al-Imam Muhammad Ibn Saud Islamic University; Saudi Arabia

*Abstract*: With the rapid development of technology and the increasing need to make quick and unique decisions from unmanaged data, there is a need to classify and analyze text data in order to extract the useful knowledge to be applied in decision-making. Text mining (TM) techniques are used to classify Arabic news into certain categories or classes. This paper aims to describe a methodology to identify word(s) tendencies in a classification process. The model constructed for classification of Arabic news will be used for this purpose.

Index Terms: Arabic news, text classification, tendency, text mining (TM)

I. INTRODUCTION

Text mining (TM), also known as textual data mining techniques, refers to the processes of extracting or discovering knowledge from a text [1]. The goal of TM is to discover unknown information which can help an individual make better decision in the future based on textual data [2].While 80% of useful information is in the form of text, such as email messages, contracts, and documents, this information exists also in the form of descriptive data written in natural language[1], [2]. The main purpose of text mining is dialed with unstructured data which data write in natural language and have more ambiguity and are not managed in a database,which makes text mining process such as clustering and classification complex .

There has been a considerable amount of text mining research done previously for the English language, but thereare fewerstudies involving the Arabic language due to the complexity associated with analyzing and classify Arabic language texts. The main challenge for analyzing and classification Arabic texts is that they have more ambiguous words (i.e.., the same word has different concepts in different contexts).For example, the word (ذهب) can refer to different concepts but is spelled the same: as a noun, it means gold, and as a verb, it means go.

In this paper, we produce a methodology usingclassification models to identify the word(s) tendencies in Arabic news text towards certain categories. We attempt to find the best classifier model that has high performance and good present the word(s) tendency. In this methodology, we tackle the following issues:

A. *Disambiguation of Words.*

Ambiguous words are words that have different meanings in different contexts. For example, the word (ذهب) can have two meanings. As a noun, it means gold, and as a verb it means "he went". (ذهب) as a noun can appear in every domain of news, but it is more likely that it will appear in Economics news and in the Beauty/Style news category. (ذهب) as a verb can be applied in news from every domain. Now, the (فرح) + (ذهب) combination can also be used in every domain, but now the occurrence level for different news can be different. It can have more tendencies toward a category like tourism and entertainment, but less tendencies toward philosophy or astronomy. (ذهب) + (فرح) + (ديزني لاند) will be strongly tendency toward tourism and entertainment and less so in religion or politics.

B. *Generate Classifiers*

We will have to estimate the classifiers that help to understand the "general tendency" of a word/group of words in training text and have better performance. Then, they will be used to calculate the "particular tendency" of word/group of words for a given text in the testing phase.

II. PROPOSED METHODOLOGY

We define the proposed methodology as having two parts. The classification part has two phases: training and testing. Phase training is used in this experiment by entering the labeled text and learned models to find the best performance model and to analyze the ambiguous words. The testing phase is used to apply the model that gains the best performance and to test it on a new text. The second part is the association rule which is used to discover the relationship between words and their tendency toward different classes. The following steps make up this methodology.

A. *Classification Part*
In this step we plan to estimate and analysis Naïve Bayes (NB), Decision Tree (W-J48) and the rule-based model such as PART and RIPPER. This part has three steps: data collection, text preprocessing and classifier generation.
 *Data set collection:* In this step, we are building an in-house corpus collected from the famous Arabic news agencies such as Aljazeera, BBC Arab, and Al-Arabiya. The corpus contains 300 documents of news distributed equally in five categories (Politics, Economics, Health, Sport, and Information Technology). Each category has 60 documents.

*Text preprocessing:* Text preprocessing is a basic step before text classification. It is applied to prepare the text by transforming it into a suitable format to be able to apply different textual data mining techniques while also removing the infrequent and worthless words [3]. This step includes series process of tokenization, filtering, stemming, and data representation.

1) *Tokenization :* The first process in text preprocessing is transforming the text from a stream of words by replacing all non-text letters such as numbers, punctuation marks (!),(?),(,),(: ),(.), (‹)…etc. and any other characters such as @,#,$,%,^,&,*,>,<….etc. to single spaces. Also, these tokenization technique results are used for further processing.
2) *Filtering:* In this step, we filter all of the characters not related to the Arabic language and the Arabic stop words that are infrequent and un-useful in text classification. Also, some Arabic words that are less than three letters are removed.
3) *Stemming (Light):* The common Arabic light stemmer applied in this research is called the Khoja stemmer, and it was reported in [4].
4) *Data representation:* In this step, we represent the textual data in matrix form to illustrate the weight of each term or feature in each text. In this research we use Term Frequency, Inverse Document Frequency (TF-IDF) as a term weighting method.

*Classifiers generation:* In this step, we plan to apply Naïve Bayes (NB), Decision Tree (W-J48) and the rule-based model such as W-PART and W-Jirp to classify the news texts and to identify the word(s) tendencies toward categories. All these models will be applied in the training phase first to learn the classification and to assess the performance of the classifier algorithms results. Therefore, we will choose the best classifier that gives the highest accuracy of classification to apply it in classification and to find the particular tendencies of word(s) for unlabeled text in the testing phase.

*B. Association Rule Mining Part*

We seek to find the relationships between words and their tendency to towards different categories and to eliminate ambiguous words. We use association rule mining for that issue. First, we work on some changes in corpus documents; we enter the label or class name into each document because we want to conclusion the categories in each association rules The minimum confidence and support are predefined as 5% for support and 90% for confidence.

III. RESULTS AND ANALYSIS

The proposed methodology in section 2 has been implemented by using the RapidMiner machine learning tool. We use the 10-flod cross-validation method provided by RapidMiner to estimate the performance of each classifier.

For the performance of each classifier, we found that the Naive Bayes (NB) achieves high accuracy than other classifiers. In, particular , NB achieves 97%, Decision tree (W-J48) achieves 75.45 %, RIPPER achieves 75.45 % and PART achieves 75.45 %.

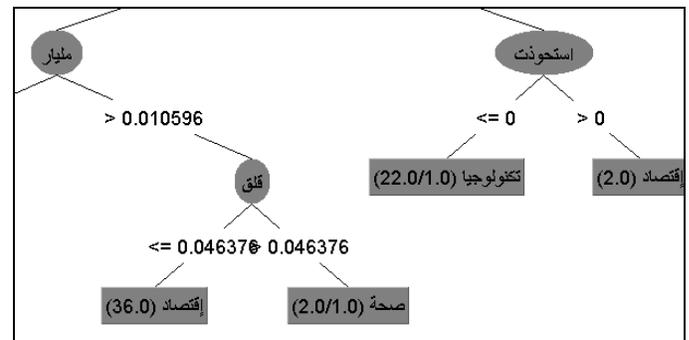

Figure 1: Partial of words classification of W-J48 classification

In regard to words' ambiguity, the decision tree (W-J48) results show some of the ambiguous words. Figure 1 shows the classification of words in the whole data set. Each (rectangular) leaf node represents the final category of words into documents, and the internal nodes (elliptic) represent the words. As we can see in Figure 1, the word قلق (anxiety) is considered to be an ambiguous word that can occur in two categories with different meanings. When the weight of قلق > 0.04 in a document, one document is classified as Health with an error of one document filled in the other category. In the صحة (Health) category, the word قلق considered to refer to the mental disease of anxiety or some things related to human health. On other hand, when the word قلق <= 0.04 is in a document, then the document given at this point is classified as إقتصاد(Economic) where 36 documents are classified as Economic news. In the إقتصاد(Economic) category, the word قلق is considered to refer to chaos and instability or fear in certain aspects associated with the economy.

IV. CONCLUSIONS AND FUTURE WORK

This paper presents a proposed methodology to identify the words tendency in Arabic news text by using the Text mining techniques. Our results show that the NB outperformed other classifier. The W-J48 shows some ambiguous words. In the future, we intend to improve our results by extracting a list of ambiguous words from Arabic news.

# Poster: An Empirical Study to Compare Sequential Forward Selection vs. Sequential Backward Elimination in Ensemble Feature Selection Setting

Faten Alrusayni*, Adel Aloraini†
* Master Student, Computer Science Department, Qassim University, Saudi Arabia, E-mail:faten3112@hotmail.com
†Head of Computer Science Department, Qassim University, Saudi Arabia, E-mail: a.oraini@qu.edu.sa

*Abstract*: The gene expression datasets represent the state of the cell in the transcriptional level and have great importance to monitor the activity and the behavior of the cell. The number of features (genes) in these datasets is very lage compared to the small sample size. Feature selection methods are essential in gene expression datasets before the analysis task using machine learning algorithms. Applying machine learning algorithms on these datasets without having feature selection steps may generate complex model and cause the overfitting problem. This paper will propose ensemble methods that combine correlation filtering ranking method with some wrapper methods that include some forms of greedy search strategies. The proposed methods will be applied on five gene expression datasets. The experimental results will show that the proposed ensemble method with sequential forward selection starting to add the highest correlated gene gives the best result in terms of prediction accuracy and survives to perform well even in p>>n problem.

Index Terms: gene expression datasets, overfitting; feature selection methods, greedy search strategies, p>>n problem.

## I. INTRODUCTION

The feature selection methods help to select a subset of features that are useful to solve the domain problem without altering the original representation of features. There are three categories of feature selection methods: filter methods, wrapper methods and embedded methods. Filter methods rank the features by assigning score according to a criterion function [1]. In the filter methods the feature selection is independent of the learning algorithm. Wrapper methods select usefulness features by evaluating the quality of subset of features using learning algorithms. To search the space of feature subsets, search methods have been used [2]. The embedded methods perform the search for features subset as part of the learning model construction [3].

In this paper, developed ensemble methods will be proposed. These ensemble methods combine correlation filtering ranking method with wrapper methods that involve some forms of greedy search strategies which are sequential forward selection (SFS) and sequential backward elimination (SBE) [4]. The proposed methods will be applied on five gene expression datasets where all samples in these datasets are cancer samples. These datasets include breast cancer, colon cancer, and three prostate cancer gene expression datasets (Table 1 summarizes these datasets). The remainder of this paper is organized as follows: section 2 summarizes search methods, section 3 discuss the results of applying our methods, and finally section 4 presents our conclusion and future work.

| Datasets | Number of genes | Number of samples |
|---|---|---|
| breast cancer dataset | 209 | 14 |
| colon cancer dataset | 98 | 5 |
| prostate cancer dataset (1) | 86 | 13 |
| prostate cancer dataset (2) | 20 | 13 |
| prostate cancer dataset (3) | 20 | 6 |

Table 1: Summary of the datasets used in the study.

## II. SEARCH METHODS

*A. Algorithm of sequential forward selection (SFS)*
1- Starts with an empty subset $F_0 = \emptyset$.
2- Finds the next informative feature $f$ that maximizes the target variable.
3- Adds this new feature to the subset $F_0 = F_0 \cup \{f\}$.
4- Evaluates the new subset $F_0$.
5- Go to step 2 (until no improvement in the prediction accuracy).

*B. Algorithm of sequential backward elimination (SBE)*
1- Starts with a subset of all features $F_0 = \{f_1, f_2, …, f_N\}$.
2- Finds the next less informative feature $f_0$ that does not improve prediction accuracy for the target variable.
3- Removes this feature from the feature subset $F_0 = F_0 - \{f_0\}$
4- Evaluates the new subset $F_0$.
5- Go to step 2 (until no improvement in the prediction accuracy).

## III. RESULTS AND DISCUSSION

In the proposed ensemble methods the subset of genes was first generated by the correlation filtering ranking method in order to find a subset of features (genes) that is highly correlated (over a threshold 0.5) with each particular gene. This subsets of highly correlated genes with each particular gene used as an input to the wrapper methods that search for the optimal subset of genes inside the subset proposed by the filtering method. Multiple linear regression (MLR) and leave-one-out cross validation (LOOCV) [5] serve as an evaluator for the proposed methods. Search methods included in the ensemble methods have been investigated using three strategies. The first two strategies work with directions according to the correlation which are: starting from the highest correlated gene and starting from the lowest correlated gene. In each direction, rank the features in the subset of highly correlated genes with each particular gene in decreasing order or increasing order, and at each time, add (or delete) the highest or lowest correlated gene to (or from) the subset of genes according to the direction used. All previous studies work with SFS and SBE in the normal direction which starts to add the most (highest) informative features (genes) in SFS and starts to remove the lowest informative features in SBE. However, in this paper, the two directions with each search method will be used. The justification is that when learning from a small sample size such as the gene expression datasets, we cannot guarantee which method will be the best. Also, in machine learning, biased to the most informative features (genes) to be in the model might not always the good way. In our situation, when we start to add the lowest correlated genes in SFS or start to delete the highest correlated gene in SBE for example, we might introduce some prediction errors but might give good prediction accuracy in the future and build most survive models. In the third strategy, the variables are not ordered. So, SFS and SBE add or remove genes (which are over a threshold) randomly as they are in the input dataset regardless to their correlation with the target variable.

The experimental results have shown that the ensemble method with SFS starting to add the highest correlated gene gave best results in terms of prediction accuracy among other strategies. This method still gives promising results even in the high dimensional dataset where the number of features (genes) far exceeds the number of samples. As figure 1 shows, SFS always propose number of features doesn't exceed the number of samples with promising prediction accuracy which indicates that this method survives with overfitting problem unlike SBE method. The results of the ensemble method with SFS starting to add the highest correlated gene was better than the results of the ensemble method with SFS starting to add the lowest correlated gene in all datasets. The results of ensemble methods with SFS without ordering variables were worse than the same method when starting to add the highest correlated gene but better than the same method when starting to add the lowest correlated gene in all datasets. SBE took more execution time than SFS as it requires more computation since SFS evaluates fewer number of features than SBE that begins with large number of features. Also, SBE gave worse results than SFS strategies except in the second prostate cancer dataset. In the second prostate cancer dataset where the number of samples gets closer to the number of genes (Table 1), no much difference between the two methods. The result of the ensemble method with SBE in the three strategies are vary as Table 2 shows.

| Strategies | breast cancer dataset | colon cancer dataset | prostate cancer dataset (1) | prostate cancer dataset (2) | prostate cancer dataset (3) |
|---|---|---|---|---|---|
| **Ensemble method –SFS (highest)** | **0.14** | **0.14** | **0.25** | **0.38** | **0.51** |
| Ensemble method – SFS (lowest) | 0.18 | 0.35 | 0.29 | 0.43 | 0.75 |
| Ensemble method – SFS (without ordering) | 0.16 | 0.31 | 0.27 | 0.41 | 0.63 |
| Ensemble method – SBE (lowest) | 69.12 | 55.65 | 3.69 | 0.38 | 1.18 |
| Ensemble method – SBE (highest) | 68.69 | 62.29 | 3.58 | 0.40 | 0.56 |
| Ensemble method – SBE (without ordering) | 72.27 | 55.12 | 2.76 | 0.41 | 0.76 |

Table 2: The overall average error result from LOOCV for all the strategies.

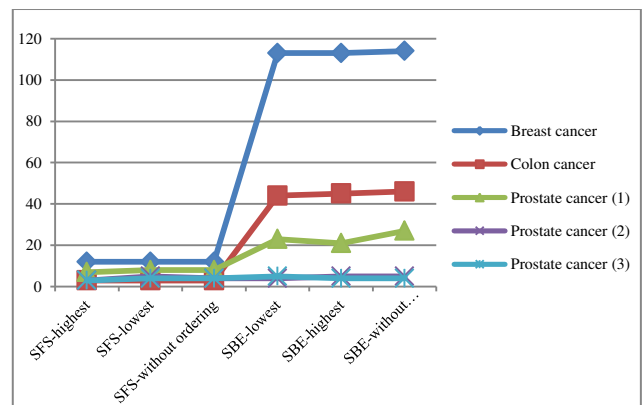

Figure 1: Maximum number of genes in the subsets proposed by the ensemble methods with the strategies.

IV. CONCLUSION AND FUTURE WORK

This paper proposed ensemble methods that integrated the correlation filtering ranking method with some wrapper methods that involve sequential forward selection (SFS) and sequential backward elimination (SBE). Multiple linear regression (MLR) and leave-one-out cross validation (LOOCV) serve as an evaluator for the proposed methods. The results of proposed ensemble method with SFS starting to add the highest correlated gene gave best result among all strategies. This method tolerated the overfitting problem and performed well in high dimensional dataset. In future, the work will look for more investigation to the ordering variables issue and seek for more methods to order variables and compare the accuracy for each of them.

# USING PREDICTION METHODS IN DATA MINING FOR DIABETES DIAGNOSIS

Arwa Al-Rofiyee, Maram Al-Nowiser, Nasebih Al-Mufadi *
Dr. Mohammed Abdullah AL-Hagery *
* Qassim University, College of Computer, Department of IT, Kingdom of Saudi Arabia
dr_alhagery@yahoo.com , nasebih@hotmail.com, arwa.alrofiyee@gmail.com, mar-26@windowslive.com

*Abstract*: Huge amounts of medical data are available currently that can leads to useful knowledge after applying powerful data analysis tools. Since diabetes been called the modern society diseases. Reliable prediction methodology to diagnose this disease will support medical professionals in this field. The research concentrates upon predictive analysis of diabetes diagnose using artificial neural network as a data mining technique. The WEKA software was employed as mining tool for diagnosing diabetes. The Pima Indian diabetes database was obtained from UCI server and used for analysis. The dataset was studied and analyzed to build effective model that predict and diagnoses the diabetes disease.

**Index Terms:** Diabetes, data mining, WEKA, artificial neural network.

## I. INTRODUCTION

Data mining is the process of selecting, exploring and modeling large amounts of data. This process has become an increasingly pervasive activity in all areas of medical science research. Data mining has resulted in the discovery of useful hidden patterns from massive databases. Data mining problems are often solved using different approaches from both computer sciences, such as multi-dimensional databases, machine learning, soft computing and data visualization; and statistics, including hypothesis testing, clustering, classification, and regression techniques [1], several research works are done in this side, but all of them are focusing on some methods of analysis, diagnosis or prediction of this disease by using different tools and methods, our work is different and concentrated mainly on the early prediction of diabetes by using WEKA and Neural networks.

The available raw medical data are widely distributed, heterogeneous in nature, and voluminous. These data need to be collected in an organized form. This collected data can be then integrated to form a hospital information system [2]. More specialized medical data mining, such as predictive medicine and analysis of DNA micro-arrays. Other data mining applications include associating the various side-effects of treatment, collecting common symptoms to aid diagnosis [3].

Diabetes is a particularly opportune disease for data mining technology for a number of reasons. First, the mountain of data is there. Second, diabetes is a common disease that costs a great deal of money, and so has attracted managers and players in the never ending quest for saving money and cost efficiency. Third, diabetes is a disease that can produce terrible complications of blindness, kidney failure, amputation, and premature cardiovascular death, so physicians and regulators would like to know how to improve outcomes as much as possible. Data mining might prove an ideal match in these circumstances, and there has been extensive work on diabetic registries for a variety of purposes. Databases have been used to query for diabetes, and to provide continuous quality improvement in diabetes care [4].

## II. RESEARCH METHOD

The research is based on three fundamentals which is the database prediction, data mining tool, and technique that will be discussed as follows:

### A. Data Set

The Pima Indian diabetes database is a collection of medical diagnostic reports of 768 examples from a population living near Phoenix, Arizona, USA. The Pima Indian Diabetes dataset available from the UCI Server and can be downloaded at www.ics.uci.edu/~mlearn/MLRepository.html. The data source uses 768 samples with two class problems to test whether the patient would test positive or negative for diabetes. All the patients in this database are Pima Indian women at least 21 years old. The database has 9 numeric variables: (1) number of times pregnant,(2) Plasma glucose concentration a 2 hours in an oral glucose tolerance test (OGTT), (3) diastolic blood pressure (mm Hg), (4) triceps skin fold thickness (mm), (5) 2-hour serum insulin (mu U/ml, (6) Body mass index (weight in kg/(height in m)^2) (BMI), (7) diabetes pedigree function, (8) age (years), (9) Class variable (0 or 1) diabetes onset within 5 years, where '1' means a positive test for diabetes and '0' is a negative test for diabetes. There are 268 cases in class '1' and 500 cases in class '0' [5]. The aim is to use the first 8 variables to predict 9.

### B. The Proposed Technique

Back-propagation is an artificial neural network learning algorithm. The technique that is applied is Multi-layer perceptron which is a classifier that uses back propagation to classify instances. The main objective of classification techniques on PIMA India diabetes dataset is to diagnose whether an individual has diabetes or not.

*C. The Tools That was Applied*

WEKA (Waikato Environment for Knowledge Analysis) is one of the most powerful data mining software tool. It is graphical user interface tool written in Java, which aims to provide a comprehensive collection of machine learning algorithms and data pre-processing tools to researchers and practitioners.

*D. Method*

The PIMA India diabetes dataset is converted to the ARFF format to be inserted in WEKA software. The WEKA Explorer provides a collection of tools, one of them is the (pre-process) tool that we used to divide the dataset to three parts: training set, testing set, and application set. The segmentation is done using the (Resample) filter in WEKA software tool, and since the dataset in the training phase effect in the learning model a 60% of the dataset used, then 20% for testing set, and 20% for application set. The WEKA (Classifier) is used to train and test the model that was build based on the training results, under the Multi-layer perceptron function. Figure 1 shows the work steps that consist of data collection and preprocessing and data analysis and knowledge generation.

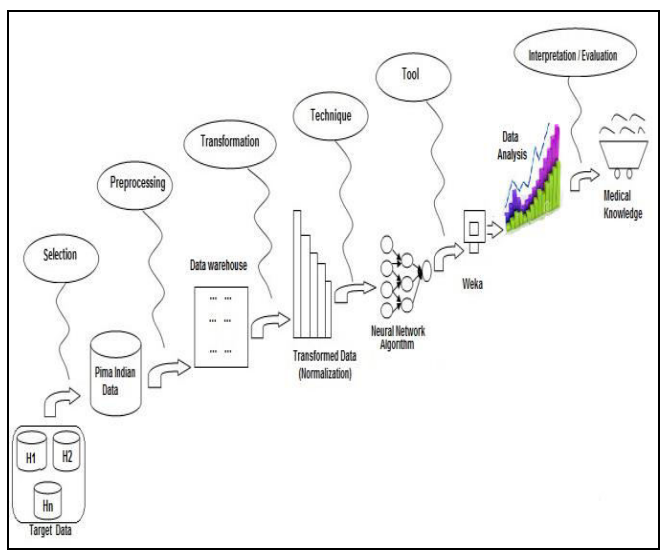

Figure 1: Work mechanism and steps

### III. EXPERIMENTS AND EXPECTED RESULTS

The results were generated using the data mining tool (WEKA) and its technique called Multilayer-Perceptron (MLP) that employed with different hidden layers value used and different training time values, which applied to the Pima Indian Diabetes database. The MLP Technique gives different prediction training value in each time the parameters of the MLP are changed, and since WEKA give diverse parameters for the MLP Technique we found after numerous experiments that the hidden layers value and training time (T.T) value parameters are the most two factors that effect in the prediction process. The highest performances of MLP in each case across the Training Set (T.S), Test Set (TE.S), and Application Set (A.S) with different hidden layers (H.L) and training time values are given in Table 1.1. The learning rate and momentum coefficient for all the training models were 0.3 and 0.2 respectively. The MLP Technique satisfied the highest correct training prediction which is 97.61% when the hidden layer parameter is set to 20 and the training time is set to 8000.

Table 1.1: Results

| Correct Prediction (%) | | | Incorrect Prediction (%) | | | H.L | T.T |
|---|---|---|---|---|---|---|---|
| T.S | TE.S | A.S | T.S | TE.S | A.S | | |
| 87.83 | 68.18 | 81.8 | 12.2 | 31.8 | 18.2 | 50 | 500 |
| 93.26 | 64.93 | 74.7 | 6.7 | 35.1 | 25.3 | 20 | 1000 |
| 97.61 | 65.58 | 73.4 | 2.4 | 34.41 | 26.6 | 20 | 8000 |

### IV. CONCLUSIONS

Through this research a practical experiences were provided to assists physician's to predict the disease earlier and to save their time and effort and also contributes to reduce the treatments costs. The research results are providing high diagnosis accuracy for this disease as well as taking advantage of using medical data sets accumulated in the UCI and employing the data mining tools and techniques for supporting the medical side, also linking the technical sciences domain with the medical services within the community.

# Methods for Climate Prediction for Saudi Arabia

Izzat Alsmadi and Mohammad Zarour

Information Systems Department
Prince Sultan University, PSU
Riyadh, Saudi Arabia
ialsmadi@cis.psu.edu.sa, mzarour@psu.edu.sa

## Abstract

Studying weather or climate historical data has been an interesting subject due to the value of information that can be utilized. Understanding hidden patterns or associations between weather attributions, locations along with prediction and forecasting are examples of how could such data be utilized.

In this paper, we described initial stages of a project that aim to conduct extensive study of weather or climate data for Saudi Arabia over a long period of time. We collected data for around 10 years including all weather stations in Saudi Arabia. Collected data represents daily weather attributes including: Temperature, humidity, pressure, wind, precipitation, etc.

We also evaluated using Time series function from R-package and WEKA to be able to come up with best methods for future climate prediction. This prediction model is evaluated using several prediction algorithms. Prediction accuracy is evaluated using some standard metrics such as ROC metrics.

### 1. General statistics

We studied all climate related data from all weather stations in KSA (29 stations) over a period of about 10 years. Data is collected for the years from 200 to 2010.

Table 1 shows general information about weather stations in KSA.

Table 1: KSA weather stations, General Information

| Station | Lat /Long | Alt/Ft | Station | Lat /Long | Alt/Ft |
|---|---|---|---|---|---|
| Abha | 18.24 / 42.65 | 6858 | Madinah | 24.55/ 39.70 | 2151 |
| Al-Ahsa | 25.29/ 49.48 | 588 | Makkah | 21.42 /39.82 | 900-1000 |
| Al-Baha | 20.29 / 41.63 | 5486 | Najran | 17.61 /44.41 | 3982 |
| Al-Jouf | 29.14 /35.40 | 2260 | Qaisumah | 28.33/ 46.12 | 1174 |
| Arar | 30.90 / 41.13 | 1813 | Qassim | 26.30 /43.77 | 2126 |
| Bisha | 19.98 / 42.62 | 3887 | Rafha | 29.62/ 43.49 | 1474 |
| Dhahran | 26.28 / 50.11 | 84 | Riyadh New | 24.95/ 46.70 | 2049 |
| Dmmam | 26.47 / 49.79 | 72 | Riyadh Old | 24.95/ 46.70 | 2049 |
| Guriat | 31.41 / 37.27 | 1672 | Sharorah | 17.46/ 47.12 | 2363 |
| Hafr Elbatte | 27.9 / 45.51 | 1355 | Sulayel | 20.46/ 45.61 | 2020 |
| Hail | 27.43/ 41.68 | 3331 | Tabuk | 28.37/ 36.62 | 2300 |
| Jeddah KAIA | 21.34/ 39.17 | 7 | Taif | 21.48/ 40.54 | 4848 |
| Jizan | 16.89/ 42.58 | 20 | Turaif | 31.69/ 38.73 | 2803 |
| Khamis | 18.3 / 42.73 | 6446 | Wejh | 26.19/ 36.47 | 66 |
|  |  |  | Yenbo | 24.14/ 38.06 | 26 |

**Time series analysis**

Time series include a sequence of data points separated uniformly in time.

We studied time series based algorithms to study evolution and weather forecasting.

Typically in such algorithms data should be homogenous. As such, we selected Amman weather station for the analysis in this section as it is complete and does not include empty values.

As a simple time series, Figure 3 shows a time series for average temperature for KSA Al-Baha station.

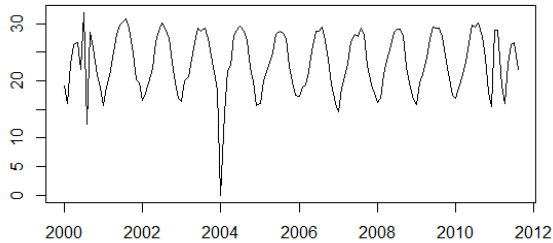

Figure 3: Al-Baha Temp. Time series

In this section, we will present using time series method to forecast weather in future based on historical data. Time series functions handle temporal data. They also acknowledge the time value within dataset and whether two records are close to each other in terms of time or not.

Time series with linear regression algorithms include a sequence of data points separated uniformly in time.

We studied time series based algorithms to study evolution and weather forecasting.

Typically in such algorithms data should be homogenous. As such, we selected Arar weather station for the analysis in this section as it is complete and does not include empty values.

Using WEKA forecasting model, we run forecasting to produce future weather attributes based on historical data. Each attribute is studied separately based on its archival data. Figure 4 shows one sample for Temperature High degree with prediction on a monthly basis for 100 future months (for Amman weather station).

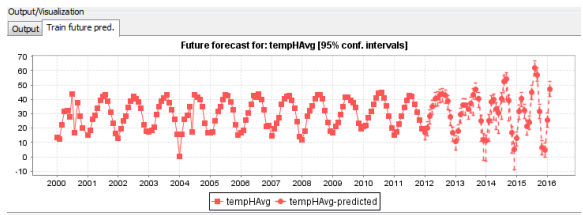

Figure 4: Temp H Time series forecasting

Figure 4 shows average temperature for each month highlighted with maximum and minimum values for months showed in dots. Figure 5 shows performance metrics for forecasting process. For both mean absolute and root mean squared error metrics, the smaller the value the better.

Figure 5: Temp forecasting performance metrics.

**Linear Regression**

Linear regression forms equations based on a goal attribute. In or case, we used precipitation as a goal class for each weather station.

```
PerAvg =

    0       * Date +
    0.0005  * dewAv1Avg +
   -0.0002  * dewL1Avg +
   -0.0001  * HumH1Avg +
   -0.0001  * HumAv1Avg +
    0.0002  * HumL1Avg +
    0.0011  * VisH1Avg +
   -0.0025  * VisAv1Avg +
    0.0013  * VisL1Avg +
   -0.0002  * WindL1Avg +
    0.001

Abha
```

```
PerAvg =

   -0.002   * tempHAvg +
    0.0064  * HumH1Avg +
   -0.0083  * HumAv1Avg +
    0.039   * VisH1Avg +
   -0.0322  * VisAv1Avg +
   -0.012

Al_Jouf
```

Attributes that show negative relation with precipitation: Humidity high, visibility average.

In case of Abha station, all attributes showed small impact on precipitation. Maximum impact was of Visibility average: -0.0025. In case of Al_Jouf station, visibility high has a high positive value in precipitation equation, next visibility average value.

**Analysis based on monthly statistics**

We noticed that time series based on daily data can be very complex to process in time series functions. We noticed also that many researchers used monthly based statistics. Hence, we averaged daily data over monthly periods for the collected data from KSA. Due to the large amount of graphs if we want to show all weather station in KSA, we will show results only for one station

We used time series functions from different applications including ( R-Package-forecast and WEKA time series). Some of the time series methods are : ARIMA, ar, HoltWinters, StructTS.

Initial analysis showed that ARIMA can be the best for our data and forecasting

Autoregressive Integrated Moving Average (ARIMA) time series algorithms or models include an explicit statistical model. It can handle irregular components of time series. This allows for non-zero autocorrelations in the irregular component.

ARIMA models are defined for stationary time series. Some preprocessing methods are used to make a time series stationary. An example of those is the function (diff) in R-Package.

We selected a fixed period of 50 months forecasting from the last observed month (August 2011). Such number can be user defined where if the number if large accuracy of prediction will be low).

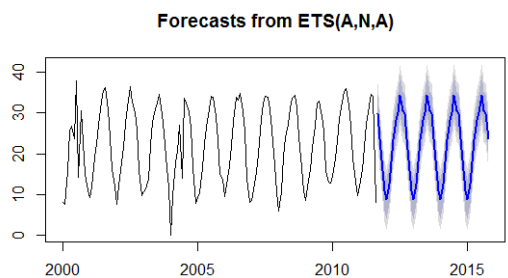

Figure 7: Arar Temp. Avg Monthly Forecast

The dark shaded region shows 80% prediction intervals. That is, each future value is expected to lie in the dark blue region with a probability of 80%. The light shaded region shows 95% prediction intervals. These prediction intervals are a very useful way of displaying the uncertainty in forecasts. In this case, the forecasts are expected to be very accurate, hence the prediction intervals are very narrow.

## Conclusion and Future Works

In this paper, we conducted several preliminary studies as part of a larger project to develop a climate prediction model for the region of Saudi Arabia. We collected a data set representing weather information from all Saudi Arabia weather stations of a period of 10 years.

We conducted preliminary studies using WEKA and R-package data mining and statistical tools. We evaluated several algorithms such as: Classification, time series, and linear regression.

We used standard performance metrics to evaluate prediction accuracy of those methods or algorithms.

We conducted further studies specially at some of the new data mining algorithms or approaches including (ensemble methods) where more than one algorithm can be used to enhance accuracy. Evaluation showed promising results. In future, we will conducted a thorough analysis for stations one by one and then try to come up with hybrid patterns that can predict future weather in Saudi Arabic based on the whole country or region by region (given the fact that Saudi Arabia has different regions based on climate classifications).

# Diabetes Early Warning System

Nouf Almutairi[1], Riyad Alshammari[2]
College of Public of Health and Health Informatics, King Saud Bin Abdulaziz University for Health Sciences, KSA
P.O. Box 22490 Riyadh 11426
Internal Mail Code 2350
[1]mutairino@ngha.med.sa, [2]alshammariri@ngha.med.sa

*Abstract*: Nowadays discovering of the knowledge from healthcare database become one of if the important applications of the data mining technique. There is a wealth of hidden information within Healthcare Information System (HIS) that can be used to support clinical decision making hence, there is a need for researches that apply the data mining in healthcare data which will help healthcare provider to predict diseases before it occurs. In this paper we present the use of knowledge discovery with diabetes data to predict the risk factors that cause that disease or increase the risk of developing diabetes in diabetic patient. The signatures/models generated from the data mining algorithms will be used in building a system that would benefit healthcare sector in providing prevention and control programs to improve the community health.

## I. INTRODUCTION

In healthcare organizations, huge amounts of data about patients and their health condition are captured by the healthcare information system. Tangible information from these data can be used in research, statistic and decisions making. Patterns and relationships that are usually found within clinical data could provide new knowledge about diseases and disease risk factors. The process of finding a relationship or a pattern within the data known as "knowledge discovery" in data or "data mining" and can be used in any field [1]. Data mining represent an efficient analytics tool used to extract and analyze patterns within the data to provide useful information. Data mining could benefit the medical field by increasing diagnostic accuracy and reducing costs [2].

Diabetes is one of the common diseases with a high prevalence across Saudi society [3]. Diabetes is a chronic diseases where human body not able to produce the insulin or use it properly. It can lead to complications such as cardiovascular, kidney failure, blindness and amputation [4]. Using knowledge discovery with diabetes to predict the risk factors that cause that disease or to predict people who are at risk of developing diabetes would help healthcare organizations to provide prevention and control programs to enhance the health of the community. Extensive data mining research has been conducted on diabetes, (Farran et al., 2013) by applying logistic regression, k-nearest neighbours (k-NN), multifactor dimensionality reduction and support vector machines as data mining algorithms on diabetes data from Kuwait Health Network, and by using only non-laboratory attributes (age, gender, height, weight and family history), they achieved an accuracy of 85% on diabetic patients and accuracy of 90% on hypertension patients [5]. Furthermore, they were able to predict 75% of diabetic patients at high risk and needed clinicians' intervention in addition to 24% of non-diabetic patients at diabetes risk who needed to control risk factors to protect them from that disease. The objective of this research paper is to use data mining algorithms, namely C4.5 (decision tree), and AdaBoost (Meta algorithm),to predict diabetes risk factors that increase the risk or development of diabetes on data collected from National Guard Health Affairs (NGHA) data. The learning model generated from the data mining algorithms will help in implementing diabetes early detection system that can be used efficiently by health care provider during treatment of patients.

## II. RESEARCH METHODOLOGY

The study will be conducted on NGHA patient's data that include patients who have visited diabetes clinic and other non-diabetic patients. Weka software [6] will be used for the data mining algorithms.

To achieve the study objective, study method consists of several phases; Data Collection, Prediction Model and Evaluation and Testing as depicted by Figure 1.

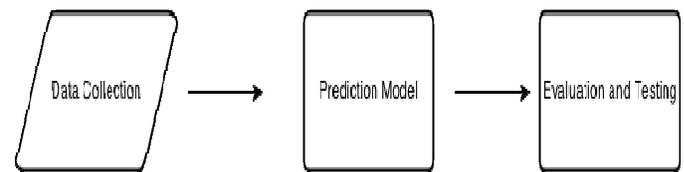

*Figure 1: Phases of the proposed project*

### A. Data Collection

Data will be collected from NGHA health information system (HIS) database. Then a cleansing process will apply to avoid unnecessary fields like missing values and abnormal values. To collect the data a request form will be sent to IT department with a clear definition of the required fields. The required fields will be as shown in Table 1.

| Variable | Definition | Type |
|---|---|---|
| Age | Number of years from patient birthday until the date of extracting the data | Continuous |
| BMI | Body mass index use patient weight and tall to determine if patient overweight, obese and etc… | Ordinal |
| Gender | Patient Gender | Nominal |
| Smoking | Dose patient smoke or not. | Nominal |
| Lab procedure | Lab procedure for each patient. | Continuous |

*Table 1*

### B. Prediction Model

Data mining algorithms C4.5 decision tree and AdaBoost (Meta algorithm) will be applied to the dataset to generate models that are able to classify patient as either diabetic or non-diabetic based on the diabetes risk factors. In this phase Weka software will be used. The AdaBoost and C4.5 algorithms are able to select the best attributes from a giving list.

### C. Evaluation and Testing

To evaluate the performance of the data mining algorithms, two metrics are going to be used, which are Recall and False Positive Rate (FPR). In this case, Recall will reflect the number of diabetic patient records correctly classified, it is calculated using:

Recall = (TP/ (TP+ FN)); whereas

FP rate will reflect the number of non-diabetic patient records incorrectly classified as diabetic patient records and is calculated using

FPR = (FP/) FP+TN)).

Commonly, the best learning algorithm is going to be selected based upon the performance of the classifiers in terms of high Recall and low False Positive Rate (FPR). Moreover, False Negative, FN, implies that diabetic patient records is classified as non-diabetic patient records, and False Positive, FP, implies that non-diabetic patient record sis classified as diabetic patient records.

### III. EXPERIMENTS AND EXPECTED RESULTS

This research proposal will employ data mining algorithms on historical data collected over a year from NGHA database. We are expecting to obtain over 180,000 records. The aims/objectives of the study is three-fold: I) To classify diabetic from no diabetic patient; ii) To identify which risk factors affected on diabetes prevalence; and iii) To identify patient's risk of developing diabetes. The aims/objectives of the study can be obtained by analyzing the signatures/models generated from AdaBoost and C4.5 algorithms. After that, the learning model generated from the data mining algorithm will be used to build a system that can predict diabetes disease in new patients; hence, providing an early warning system to patients and healthcare providers.

### IV. CONCLUSIONS

Healthcare data is going to be analyzed to discover hidden information and extract knowledge to improve the health status of the diabetic patients. For this study we expected to build an early warning system based on the learning model generated from applying data mining algorithms on data collected from NGHA database over a period of one year. Such a system can serve as an assistant tool for physicians and nurses to make better clinical decisions and also it can be utilized for patient protection and control plan.

# Finding a relationship between trending Arabic news and social media content using data mining techniques

Nourh Abdulaziz, Rsheed*, Dr. Muhammad Badruddin Khan †
* College of Computer and Information Sciences, Imam University, Saudi Arabia
† College of Computer and Information Sciences, Imam University, Saudi Arabia

*Abstract*— **Predicting the spread of news articles via social media is a challenging task. The trending news in the cyber world is subject to the attention of most web users in the internet world. Websites of news channels like the BBC, CNN, etc. are the sources of emerging news. The trending news is expected to make an impact on the social media world, where it can be shared and discussed or commented on. Twitter is a venue that has attracted hundreds of millions of people who share their views. Trending news is also discussed in different tweets. In this paper, the phenomenon of spreading trending news is discussed. Each news article has certain set of features that plays its role in the popularity of trending news in the Twitter world. We describe our approach to using these features to predict the popularity of news articles using data mining techniques.**

*Index Terms*— prediction, data mining, news, Twitter, popularity.

## I. Introduction

The Wikipedia online encyclopedia describes news articles in the following words. "A **news article** discusses current or recent news of either general interest (i.e. daily newspapers) or of a specific topic (i.e. political or trade news magazines, club newsletters, or technology news websites)". In this era of information, most news sources (like newspapers and TV channels) offer online news via their websites. The online news websites enable internet users to see/watch the news on their internet-connected machines, which may be desktops, laptops, or handheld smart phones. There are also websites that aggregate news from different sources, thus helping internet users to save time and pick and choose the news from any source in which they are interested.

As for any language, Arabic news websites attract an Arabic-speaking population. Arabic newspapers have their respective websites that are regularly updated. Arabic news channels (such as Aljazeera, Arabic CNN, and Arabic BCC) not only capture the attention of millions of viewers via TV, but also have a very strong and effective presence in the cyber world through their websites. In recent years, the popularity of trending news has been enhanced through social networks like Twitter. A single tweet with a link to news can cause thousands of clicks on a website carrying a particular news article. The phenomenon of tweeting and re-tweeting not only extends the reach of the news, but also causes the news to "live" for more time. The difference in the popularity of news varies and is based on features of the news articles.

Prediction of the extent of the spread of news articles via social media is a challenging task. In this work, an approach to finding a relationship between the "trending Arabic news article features" and the "level of coverage of articles in the social media contents (tweets)" is discussed. This study is significant because it will help to discover/formulate models that will predict the popularity of news articles before they are published. The main objective is to discover whether it is possible to make reasonable forecast of the spread of news article based on the news article's features. We will use data mining techniques to achieve our objective. In this paper, we propose models for internal and external features to predict the popularity of news articles.

## II. Approach Proposed

The analytical approach of this research is based on classifying news articles into three categories according to Twitter popularity (the number of tweets for each article), in a range of high, medium, or low popularity.

In our research, there are three models, one for the internal, one for the external features and one for internal and external of the articles. These three models rely on several main steps. First, the news articles and the number of tweets need to be acquired online during the data acquisition process. Then, the article and identifiable features for each article need to be labeled for training purposes and data testing. External features refer to predicted popularity according to the features of the article, such as the article's category, news source, host website, language (subjective/objective), and presence of famous names. The external feature model focuses on analyzing data. The internal feature model predicts popularity based on the words of the article; it focuses on analyzing the text. Third model predicted popularity according to the features of the article and based on the words of the article.

### III. WORK FLOW

To complete this model we took three steps:

1. *Data Collection*
   - News articles published earlier in 2014 on three major online news websites: BBC World Service Arabic (BBC), Al Jazeera Arabic (AJA), and Cable News Network Arabic (CNN) were collected. The articles were downloaded periodically from the homepage of each of these websites; only the trending news was downloaded.
   - Features of downloaded articles were saved along with the title and content of the news.
   - Twitter's Topsy was used periodically for the purpose of tweets collection. How many times the news link was tweeted or re-tweeted was recorded.
   - The collected data was saved in an Excel file.

2. *News Labeled by the Number of Tweets*

   A collection of manually labeled documents was created. The labels were determined by the number of tweets for each article (high popularity, medium popularity, and low popularity).

3. *Document prepossessing*

Preprocessing consists of two steps:
   - Tokenization: The text of the documents is separated into individual words. The news article (text of document) is split into a sequence of tokens.
   - Stopwords: This process filters Arabic stopwords from a document by removing every token which equals a stopword, such as a preposition.

4. *Extract Features*

   According to previous studies [1], there are no identical approaches for feature extraction for Arabic text classification. We applied bag of words, stemming, n-grams and stemming and n-grams.

5. *Classifier Learning*

   Three learning methods were applied : Decision Tree, Naïve Bayes and K-NN. Comparisons were made between these methods' performance to identify relationships between the external features of news articles and their Twitter popularity. After the news articles were labeled, they were randomly divided into two sets: a training set and a test set. Typically, the division places approximately 75% of the documents in the training set and the remaining 25% in the test set. The training set is used to train the classifier, while the test set and a manually labeled set are used to evaluate system performance.

The same experiment was applied for the external and internal features in Rapidminer[1], in preparation for data/text mining and importing the data and text. The Rapidminer application is an environment for machine learning and data mining processes; its focus is on predictive analysis.

### IV. RESULTS

The results of the experiment on the external features are that the decision tree delivers higher performance than KNN and NB. It has good performance at 93.30%. Website, source, and category features are significant for predicting the popularity of news articles. Subjectivity of the language and famous names are not significant features for predicting the popularity of news articles in Twitter. The experiment on the internal features yielded no significant predictors of the popularity of news articles in Twitter The results internal and external feature in same experiment KNN delivers higher performance with different methods of external features than Decision Tree and NB. It has good performance at 92.00%. Figure 1. Illustrate the performance for each method in different method of external features.

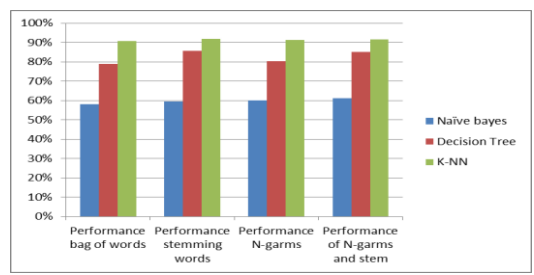

*Figure 1: Summary of performance for each method*

### V. CONCLUSION

In this work, we have discussed an analytical approach that enables a person to predict the popularity of news articles in the Twitter system, based on news article features. Category, website and source of articles at external features significantly to predict popularity in twitter. Internal feature that contain Arabic text do not significant to predict popularity news articles in the Twitter system.

---

[1] http://rapidminer.com/

# Understanding the Content of Arabic Tweets by Data and Text Mining Techniques

Rehab Nasser, Al-Wehaibi*, Muhammad Badruddin, Khan†
* College of Computer and Information Science, Al-Imam University, Saudi Arabia
† College of Computer and Information Science, Al-Imam University; Saudi Arabia

*Abstract*—In recent years, the web of social media has attracted millions of individuals to communicate online. As in the rest of the world, users in Arab countries engage in social media applications for interacting and posting information, opinions, and ideas. Because these media have become an integral part of life, analyzing the enormous amount of data available in social media applications is a valuable endeavor. In this paper, we will present a roadmap for understanding Arabic tweets through two main objectives: 1) to predict tweet popularity in the Arab world and 2) to analyze the use of Arabic proverbs in tweets. The internal and external features of the content of tweets were used as input for data mining techniques to understand the phenomenon of tweets in the Arab world. Our experiments consisted of application of classifier like Ripper (Rule Base), Naïve bayes, and Decision Tree on Arabic tweets.

**Index Terms: Social media, Twitter, Naïve Bayes, data mining, text mining, Arabic language**

## I. INTRODUCTION

In this paper, we emphasize two main points to understand the nature of Arabic tweets based on content. First, we look at why some tweets spread more quickly and broadly than others. To answer this question, we examined external features using data mining techniques. We also examined internal features (text of the tweets) using text mining techniques to understand the relationship between content of tweets and their spread level. Second, we investigated the usage of Arabic proverbs in Arabic tweets. The idea was to check in which context Arabic proverbs are used in tweets and what internal features cause these proverbs to be used in different context.

Text mining in the Arabic language is a difficult task because the language is highly complex, with grammatical rules that are considered more complicated than the grammar rules of many other languages [1][2].

Since our goals are to understand the tweet content in the Arab world, we focused on informal Arabic language. This is primarily because the majority of Twitter users in the Arab world compose their tweets in informal Arabic, particularly when incorporating Arabic proverbs to make a point. Informal Arabic language presents other challenges, such as its tendency to repeat one letter in some words. In addition, some tweets included spelling errors.

This study provided a valuable roadmap for analyzing Arabic language content in the Twitter platform from the Arab world and presents useful results for future research.

The remainder of the paper is organized as follows: Section 2 presents the data collection method; section 3 shows the methodology for the study; section 4 illustrates the analysis of the results; and section 5 presents the research conclusions.

## II. DATA

### A. Tweet popularity

Data for the tweet popularity analysis was collected from the top ten most active personal accounts in Saudi Arabia. We focused on Saudi Arabia because 32% of active Twitter users in the world are from this nation according to a 2013 study from the PeerReach website[1]. Our sample consisted of 600 tweets. We examined the tweets based with respect to two dimensions.1) External features 2) Internal features:

*External features:*
1- Does the tweet contain a URL?
2- Does the tweet contain a hashtag?
3- If it contain hashtags, how many are present?
4- How many text characters are in the tweet (tweet size)?
5- Tweet category (sports, religious, political, and ideational).
6- Number of retweets (#ofRT).
7- Level of popularity based on the number of retweets. If #ofRT >= 1000, then tweet has high popularity; if #ofRT >= 500 and < 1000, then tweet popularity is medium; if #ofRT < 500 then tweet has low popularity.

*Internal feature*
We examined the text of the tweet to determine which word made the tweet more popular than the others and vice versa.

### B. Arabic proverbs

Ten Arabic proverbs were included, and 44 tweets were collected for each. This generated a set of 440 tweets containing Arabic proverbs. We examined the internal features to determine which word qualified the Arabic proverb for inclusion in a specific category.

---

[1]http://blog.peerreach.com/2013/11/4-ways-how-Twitter-can-keep-growing/

### III. METHODOLOGY

Our methodology began with data collection from the Topsy search engine. The collected data was then entered into Excel. The Excel sheet was imported into RapidMiner tool for analysis. The research methodology illustrated in Figure 1.

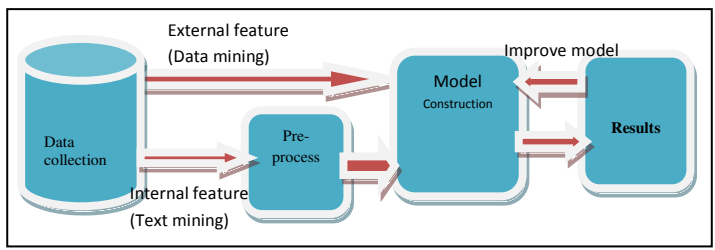

Figure 1: Research methodology

#### A. Pre-process phase

Here, we cleaned the data set text, making it suitable for use in the model construction phase. This stage included the following steps:
- Tokenize: Tweet text was split into a sequence of tokens.
- Tokenize 2: By mode, "regular expression" was used to remove English words.
- Filter step word: Arabic stop words, such as "من", which means "from" in English, were removed.
- Replace token: Spelling errors, such as "ممتااااز" spelled as "ممتاز", were corrected. Arabic characters were normalize for unification; for example, "أ،إ" was normalized by replacing it with "ا".
- Light stem: Suffixes and prefixes were removed.
- Filter token by Length: Useless words < 3 characters were removed.

#### B. Model construction phase

Two study models were constructed in this phase. The first series of experiments was planned to construct model to predict tweet popularity, and the second series of experiments was planned to classify Arabic proverbs usage context based on content or internal features of tweets. The model construction phase contained the following of Rapid Miner (a popular tool for data mining purpose):
- Process document from data: Steps were pre-processed inside this operator to prepare text and data before inserting external and internal features into the model.
- Set role: The labels for each model were defined. The predict-popularity model was labeled "Popularity" with three class values (high, medium, and low) according to the seventh step we mentioned under Eternal Feature in section 2. The Arabic proverbs classification model was labeled "Category" with four class values (sport, religious, political, and ideational).
- X-validation (cross validation): Data was spilt and entered into the model in two groups: training and testing.

We first examined the external features in the initial model and the internal features of the tweets in the two models. We used three algorithms for this purpose: Naïve Based (NB), Decision Tree, and Rule Base (Ripper) which called in RapidMiner program (W-JRip).

#### C. Results phase

In this phase, we obtained the results for each model. This result explained average performance, class recall, and class precision. If the result was not acceptable, we returned it to the model and improved the process by using feature selection or making other improvements to the pre-process phase and so on.

### IV. PARTIAL RESULTS AND DISCUSSION

We will discuss in this section part of our results.

After series of experiments we constructed the first model that predict level of popularity for Arabic tweets. Based on the external feature we obtained the classification model generated by Decision Tree was the best result until now with accuracy 87.92%. In other experiment we construct the model based on both external and internal features and we got the best result from the Ripper process (Rule Base) that presents some rules to predict the level of Arabic tweet popularity in a good manner with accuracy 86.45%.

There are two external features that play main role in prediction the level of tweets popularity which are tweet category and tweet size. Also, we notice that when the tweet content related to sport and religious categories will has more chance to be high in popularity. Also, when the tweet category was (Ideational and political) then the Hashtag as an external feature will affect the tweet popularity level; if tweet contains hashtag will be high or medium in popularity level but if tweet doesn't contain hashtag will has low chance in popularity.

### V. CONCLUSIONS

This paper presents two models to analyze Twitter in the Arab world concerning tweet popularity prediction and the classification of common Arabic proverbs in tweet content. These two models were designed to increase the social understanding of Arabic tweets. We examined three processes (Naïve Bayes, Decision Tree, and Ripper) by using RapidMiner tool to identify the training model that produced the best tweet classification results. We chose the best model based on the training model classification results as well as the accuracy of each process. As future work we intend to increase our data set. In addition, we plan to test other classification processes, such as K-NN and SVM, to make a comparison between multiple classification processes.

# ShaMoCloud: A Mobile-Cloud Framework for Mining Video Data

Fahad Almudarra, Basit Qureshi*
Email: {fahad.almudarra@gmail.com, qureshi@cis.psu.edu.sa}
*Prince Megrin Data Mining Center
College of Computer and Information Science,
Prince Sultan University, Saudi Arabia

*Abstract*: Mobile cloud computing is the current trend for sharing images and videos. Researchers have estimated that video data traffic will contribute up to 66% of mobile data traffic in the coming years. , This massive increase in video data traffic opens new challenges for video data mining and pattern detection in various applications including surveillance, security etc. In this paper we address this challenge by suggesting a mobile-cloud framework for mining video data. In our initial implementation we tried to apply "Video annotation" or "Tagging" concepts in order to attach keywords to specific video and cluster them into groups.

**Index Terms: Video Data Mining, Video information retrieval, Mobile Cloud Computation.**

## I. INTRODUCTION

Integrating mobile applications with cloud environments have recently become a popular approach in the industry. In recent years there is a trend in migrating popular applications to the cloud with social networking applications such as Facebook, twitter etc. amongst many others. Furthermore, many software giants have been offering their software and services in cloud such as Google-docs, Gmail, MS-Office 360 etc. Microsoft Azure, Google App Engine, Amazon EC2 and recently IBM CloudOne have been providing Software-as-a-Service (SaaS) for developers to build applications using corporate infrastructure.

As reported in [3] it is expected that the video data traffic will generate 66% of mobile data traffic. Researchers expect the video data traffic to increase many folds in the future. In light of these predictions, enabling resource-constrained mobile devices in order to handle the computation-intensive applications is one of the main benefits of the mobile cloud computing [4]. Such environment would allow users to share videos captured or stored on local mobile devices to online cloud service.

Since the video data traffic will increase and the searching features for a massive and huge data will be more complicated, the data mining is proposed solution to improve searching behaviors and user experience. Data mining is a process of extracting previously unknown knowledge and detecting the interesting patterns from a massive set of data [1]. Video data mining that deals with the extraction of implicit knowledge, video data relationships

"ShaMoCloud" is mobile application integrated with cloud service that will be used to apply proposed framework. The proposed framework applies "Video annotation" or "Tagging" concepts in order to attach keywords to specific video and cluster then into groups. We intend to show this concept will facilitate the searching mechanisms for users.

## II. PROPOSED SOLUTION

We intend to develop a mobile cloud framework that will allow:
*Content sharing*
The user can share content with his authorized friends within the cloud. This service is related with Content Management service.
*Content management*
User can manage video multimedia content in cloud from his mobile device directly. That will allow user to make backup of his content and grant privileges and permissions to his friends to sharing this content.

The mobile application will be developed with cloud computing service and Cloud computing provider allows us to use and share the combination of software and hardware as service. **Fig 1**

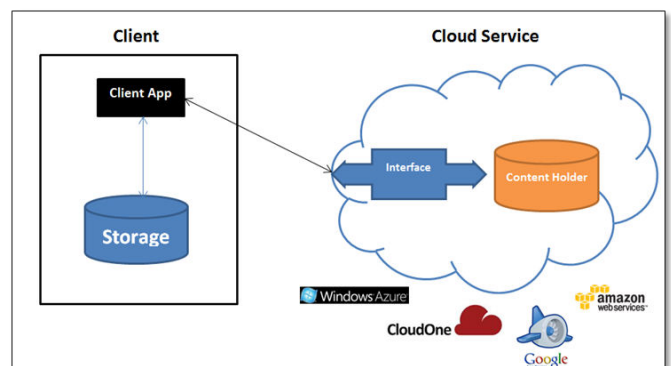

**Figure 1 Cloud Computing Service Architecture**

We will build a mobile content sharing application based on Android platform. The applications would be developed and hosted by:

A. *Public Cloud Service Provider:*

A popular cloud service provider which may include Amazon EC2, Google App Engine, Windows Azure or CloudOne and it would be accessed on mobile devices as shown in Figure 1.

B. *Private cloud service*

We are planning to develop and host the application in private/internal network and servers we plan to use Open Stack SDK as cloud computing environment. Open Stack SDK would be installed on a server which is going to emulate a large scale data center.

III. SYSTEM ARCHITECTURE

*(a) Mobile App Architecture*

Component diagram in **Figure 2** shows the architecture from mobile application side. It consists of the following components:

1. *Content Management*: This component is the main component from mobile side. It presents the main functionalities of this project. Like upload, share and delete videos from cloud service side.
2. *User credential:* it contains user credentials for the video owner.
3. *Encryption:* This component used for encrypt the user credential through connecting with cloud.
4. *Mobile Device Storage DB:* This database is the mobile device database itself. It is needed to retrieve videos from the device.
5. *User Directory DB:* It is database used for storing and returning the user credentials details.
6. *Mobile Interface:* it is an interface contains the main agreement points between mobile and cloud service; what is needed and required from another.
7. *Amazon EC2 provider:* it presents the cloud service side.

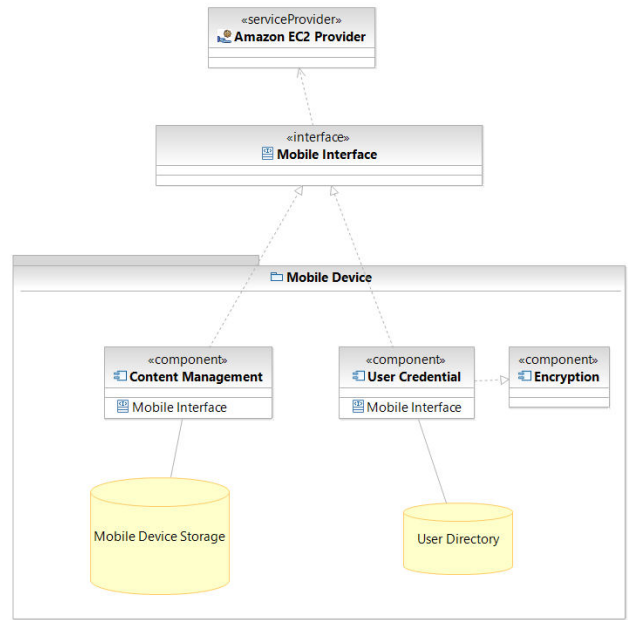

**Figure 2 Mobile Application Component Diagram**

*(b) Cloud Service Architecture*

Component diagram in **Figure 3** shows the architecture from Cloud service side. It consists of the following components:
1. *Content Management:* This component is the main component from mobile side. It presents the main functionalities of this project; like upload, share and delete videos from cloud service side. Here is from cloud service side, it will contain the detailed commands like inserting, delete and granting privilege of sharing.
2. *Videos DB:* it contains videos media.
3. *User credential DB:* it contains user credentials for the video owner.
4. *Users DB:* It is contains the granted user to have the accessibilities to videos and be shared.
5. *Amazon EC2 Interface:* it is an interface contains the main agreement points between mobile and cloud service; what is needed and required from another.
6. *Mobile Application:* it presents the mobile service side.

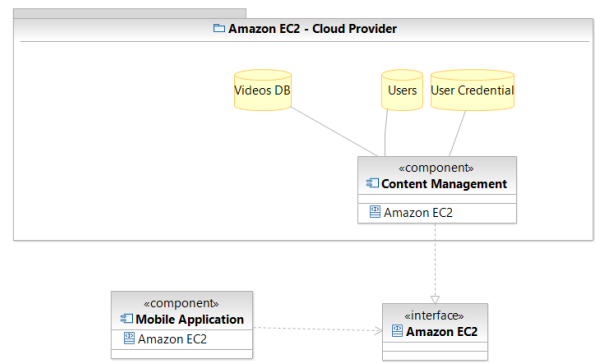

**Figure 3 Cloud Service Component Diagram**

## IV. VIDEO ANNOTATION AND TAGGING

In our proposed framework, we consider in the design step to adopt a famous concept used in the data mining field which is "Video annotation" or "Tagging"..

*Annotation involves attaching keywords from the users along with commentaries produced by experts and those obtained from a body of the existing texts. The text embedded in the video (closed caption) is a powerful keyword resource in building the video annotation and retrieval system enabling text-based querying, video retrieval and content summarization"* [1].

User will be allowed to add text tags to videos captured. These videos would be uploaded in the ShaMoCloud platform. Videos DB as shown in figure 3, contains video stores by user, user information and tags associated with each video. This information is essential to a user for their video query.

This value accepts multi value or list. From Video DB, we can extract and retrieve video based on specific tag(s). Then, Video Mining done by clustering the tag(s) into similar groups. Ultimately, compress the clustered video in the storage as shown in **figure 4**. This will improve user experience and search behavior for the user.

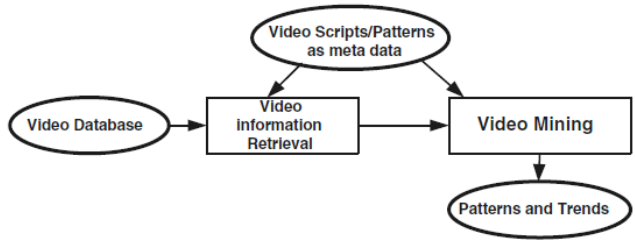

**Figure 4 Video Mining**

## V. CONCLUSIONS AND FUTURE WORK

ShaMoCloud is mobile application integrated with cloud service. It is developed to provide storing, sharing videos in cloud. This paper presented "Video annotation" or "Tagging" technique which used in video data mining through MCC We embedded the discussed Video Mining framework to our framework.
In future will study the current framework and analyze it in order to provide practical and experimental study to improve this framework. Also, we may apply a more mature framework proposed by experts in video data mining since the integrating mobile cloud computing and video data mining need experience exchange in both fields.

# Poster: Learning Sentiment from a Dynamic, Short and Unstructured User-Generating Content.

Salha, al Osaimi , Muhammad Badruddin, Khan
Dept. of Information Systems Imam Muhammad ibn Saud Islamic University, KSA

*Abstract*—This paper discusses the issues related to sentiment analysis of user-generated content in the world of social media. Although the focus is on the content generated in Arabic Language, the core concepts and most of the techniques are generic and language-independent. Millions of Twitter's tweets with modern Arabic content provide a challenging opportunity to understand the emotions of their producers. This paper investigates how the sentiments present in the Arabic tweets can be captured using text mining techniques.

**Index Terms:** Twitter, User-generated content, Sentiment analysis, Opinion Mining (OM).

## I. INTRODUCTION

Just within few years, world has witnessed its population communicating with each other in new way in the new medium. There are wide numbers of users that use the social networks. They use social media like Twitter, Facebook, MySpace to share various kinds of resources, express their opinions, thoughts, and messages in real time, thus resulting in increase of amount of electronic content that is generated by users. This electronic content is the input for Text mining techniques that produce useful information from the data.

The Arabic language is spoken by more than 330 million people as a native language. Arabic is official language of Islam and was selected to be the language of Holy Quran and language of the last Prophet. Hence Muslims living throughout the world feel affiliation with Arabic language.

Practically, social media sites play an important role in an internet users' life. One of these sites is the Twitter which has enormously popularity with more than 50 million users logging in daily and billions of tweets are sent each month. According to value of Arabic language and how much the social media affect the society, this work proposes to serve the Arabic language in one of important social media platforms. In this work we focus on analysis of informal Arabic tweets in phrase or sentence level [1].

Sentiment Analysis aims to identify emotions, opinions, attitudes of speaker/writer of the content with respect to some issue. The sentiment analysis has different names such as: opinion mining (OM), subjectivity analysis or sentiment orientation [2]. There are numerous applications for sentiment analysis of English content, whereas field of Arabic content sentiment analysis is still far away from mature level and limited numbers of tools are available.

Since most of users use informal Arabic in the world of social media, the task of sentiment analysis becomes more sophisticated. Different Arabic Dialects are another challenge [3]. One of the main challenges is the limited researches that focus in the Arabic sentiments analysis. This motivate us to focus on the problems that exist in informal Arabic sentiment analysis thus paving the way for current and future researchers to participate more actively in this field.

In twitter the users share short pieces of information known as "tweets" (limited to 140 characters). Twitter sentiment analysis is not an easy task because a tweet can contain a considerable amount of information in very compressed form. Moreover, the tweets have specific unique characteristics as compared to other domains like review website, which add new challenges for sentiment analysis [4]. This fact also motivated us to work on tweets rather than content of other social media like Facebook. Also, the sentiment analysis on twitter can be extended to serve varieties of application that serve the community such as business or political applications.

The rest of the paper is organized as follows: Section II describes the methodology of our work; section III shows the initial results followed by conclusions in section IV.

## II. PROPOSED METHODOLOGY

We would like to improve the performance measures of informal Arabic language at phrase and sentence-level sentiment analysis by proposing two approaches. First approach is to study the emotions icons in the informal user-generated texts to discover the sentiment analysis. Also we try to explore the emotions icons that are only being used in Arabic text.in our first approach; we will also try to find sentiments in Arabic tweets without emotion icons in future. In the second approach, we plan to include semantics of words for sentiment discovering process. We are now doing experiments under first approach. The first approach requires us to follow the following steps. Figure1.1 illustrate the sentiment analysis model for informal Arabic with emotion icons

### A. Collection of Data (the Arabic tweets)

In order to collect a big corpus of data with Arabic tweets, we automate collection by using twitter's API library.

In order to address ambiguity problems, we will make separate word list carrying ambiguous words so that tweets with such tendencies can be sorted out. This task will be performed when we will start work under second approach.

### B. Data Preparation

First, after we collect tweets that must contain at least one emotion icons. We assigned class (Sentiment) to each tweet manually. Hence in this step two raters read each tweet and assign its respective class (positive, negative and neutral). They had a high degree of agreement in their classification of the tweets, and for those tweets that disagreed; a third rater was used to determine its final sentiment

In fact, the texts that is produced from the users in social media is mainly consider as unstructured and noise nature. This is due to the informal Arabic nature which is unstructured and with no grammar standardization. Also, the spelling mistakes and missing punctuation produce noise. That is why the preprocessing step is very important to prepare data to classification step. There are five steps for the preprocessing data:

#### 1- Normalization:

The normalizing process is manipulating the Arabic text to produce consistent form, by converting all the various forms of a word to a common form. For example the word "الله" can have many different forms like "اللـه", "اللــه", etc. All these forms cause the single word to be considered as three different words. So, we need to transform them to a single form. In this step we also do several tasks such as removing diacritics from the letters, removing „ء" (Hamza).

#### 2- Convert the emotion icons:

The convert emotion icon is process to convert each icon to its meaning because when we clean the text from English word and special characters all those emotion icons were removed. In order to preserve the emotion icons, we gave special meaningful name to each emotion icon. Table1.1 show examples of the emotion icons conversion step.

Table 1.2 examples of the emotion icons convert steps

| Emotion icons | Tweets | Tweets after convert the icons |
|---|---|---|
| ^_^ | صور] عيد ميلاد سعيد لـ جيجونغ من [ ^_^ المعجبين في البيرو | صور] عيد ميلاد سعيد ل جيجونغ من [ المعجبين في البيرو  رمزخجول |
| ♥ ):' | مع السلامه :( ♥ | مع السلامه رمزحزين   رمزقلب |

#### 3- Clean the dataset

The clean process is used to remove all user-names, pictures, hash tags, URLs and all non-Arabic words from the tweets to manipulate them easily.

#### 4- Remove stop word

There is not specific list of stop words for Arabic. Depending on the type of the application, different lists are being used. Furthermore, there is no given stop word list for informal Arabic language. We have to build this stop word list.

### C. Building of Model

We used "Rapid Miner" to build model to classify the tweets with correct sentiments. The training data with assignment of sentiments to tweets was the input of the process. We plan to use text mining techniques in future to disambiguate ambiguous word automatically and then using results in sentiment analysis of tweets

### D. Model Description and Validation

After the development of model, we validated the model using the cross validation technique. If the model is accurate enough, then we will try to analyze the model and try to find out which "Arabic words" are the key players in the process of assigning sentiments to a tweet.

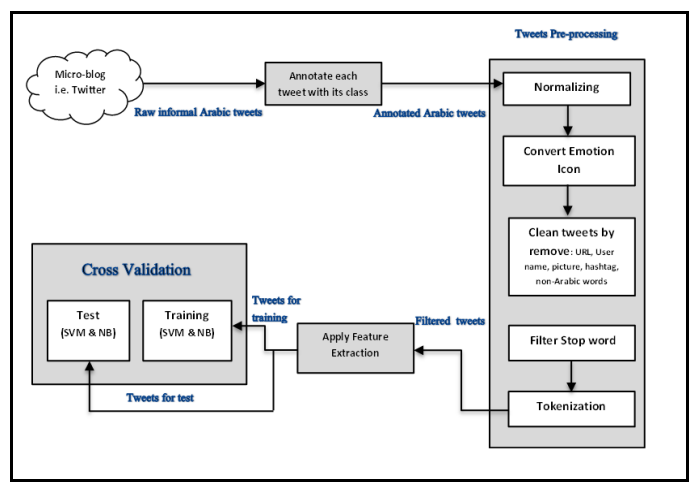

Figure 1.1 Sentiment analysis model for informal Arabic tweets contain emotion icons.

## III. THE EXPERIMENT RESULTS

The main result is there were no specific rules when the users using the emotion icons in informal Arabic tweet. For example, we found many tweets which has negative opinion and contain a positive emotion icons such a smile face. Also we found some specific Arabic emotion icons we call it Islamic icons such as: hala ☾, Masjid ⌒☽ ⋏, Verse icon ✸,﴿١﴾. Also there are other Arabic emotion icons that used Arabic's letters and symbols like anger face (ب_ب) and crying face (٤_٤) . Finally, we found there is a limitation the number of Arabic sentiment lexicons. Furthermore, there is no specific list for informal Arabic stop words.

## IV. CONCLUSION

Research in sentiment analysis for the Arabic language has been very limited as compared to other languages like English. This paper describes the issues related to sentiment analysis of user-generated content in social media (Twitter ecosystem) that are written in Arabic Language. We discussed the approach that we plan to take to tackle this issue. We collected data and perform few experiments on tweets with emotion icons. We have discovered that the emotion icon that seems to be positive in nature, also are used in tweets with negative sentiments and thus it is difficult to build a sentiments classification model based on just emotion icons. In future, we plan to perform rigorous experimentations on all words of tweets to develop a reliable classification model for sentiment analysis

# Using Machine Learning of Graphical Models for a better understanding of cancer development

Bayan, Almkhelfy*, Dr. Adel Aloraini†
* Information Technology department, Qassim University, Saudi Arabia
† Computer Science department, Qassim University, Saudi Arabia

*Abstract:* The work in this paper aims to show how the genes interact with each other when the breast cancer is developed. Several Machine Learning of Graphical Models methods such as Co-expression network, Bayesian network and Dependency network are applied in gene expression dataset to find interested interactions between genes. In this work, we compared the performance of these methods, using Leave-One-Out Cross Validation (LOOCV) in terms of prediction accuracy. The goal of this work is to find out which one of our methods gives more reliable result to learn gene-regulatory networks when cancer is developed.

**Index Terms:** Machine Learning of Graphical Models, gene expression, gene-regulatory networks.

## I. INTRODUCTION

Recently, breast cancer considered to be one of settled problems especially in women community. Hence, it is important to understand how the genes interact with each other, when the cancer develops in order to develop drug strategies. Interestingly, machine learning methods can offer the ability to help in monitoring the change of gene expression when cancer is developed. In this work, we will demonstrate the applications of different Machine Learning of Graphical Models methods in gene expression for breast cancer samples. Then, we will evaluate the performance of these methods using LOOCV to find out which method considered more reliable to learn from gene expression.

## II. METHODS AND RESULTS

### A. Learning Co-expression network

To learn a Co-expression network, a correlation coefficient method (1) with a threshold equals to 0.5 will be used to control the connectivity between genes.

$$r = \frac{\sum_{i=1}^{n} Y_i(X_i - \bar{X})}{[\sum_{i=1}^{n}(X_i - \bar{X})^2 \sum_{i=1}^{n}(Y_i - \bar{Y})^2]^{1/2}} \quad (1)$$

where $n$ is the number of samples, $\bar{X}$ is the mean of a particular gene expression values($X$), and $\bar{Y}$ is the mean of a another particular gene expression values($Y$). If correlation coefficient($r$) is larger than or equals the predefined threshold, then the genes pair are connected in the graph by line. The range of the $r$ values is between [-1, 1], when r = 0 then no linear correlation between variables. Figure 1 shows the result of applying Co-expression network when learn from breast cancer dataset with a threshold >=0.5.

The value of the effectiveness between two genes in the graph is controlled by the thickness of the line. The resultant Co-expression network represents regulatory relationship between genes. The regulatory relationships between genes in the resultant Co-expression network express the Dependency between genes. One drawback of using Co-expression networks is that we cannot determine the cause and effect relationships. Therefore, we will experiment with Bayesian networks to find out the directionality of the relationships.

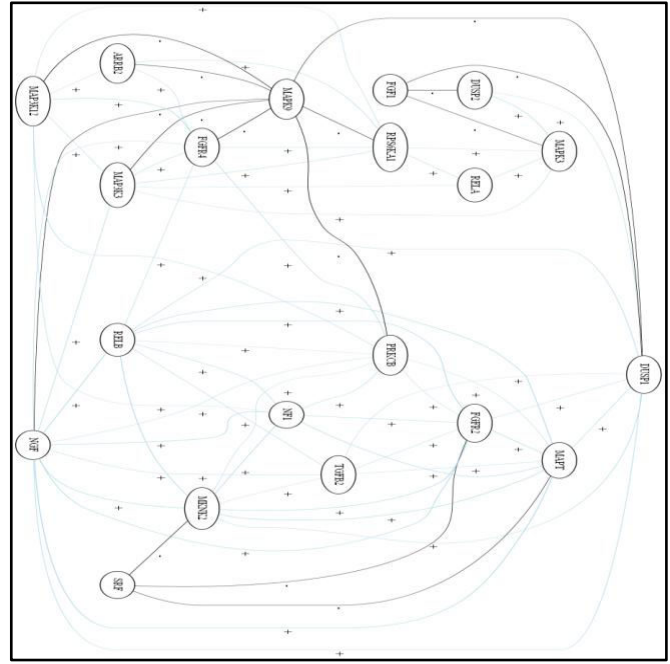

Figure 1: The model resultant from Co-expression network learning method

### B. Learning Bayesian networks(BNs)

To learn a Bayesian network, we used simple linear regression(2) to select the best parent for each gene from all possible parents that has the smallest residual sum of squared errors and constructs the Bayesian model from the dataset accordingly.

$$Y = \beta_0 + \beta_1 x + \epsilon \quad (2)$$

Where $\beta_0$ is the intercept, $\beta_1$ are coefficient regressions and $\epsilon$ is a random error. Figure 2 shows the result of applying Bayesian network method using simple linear regression.

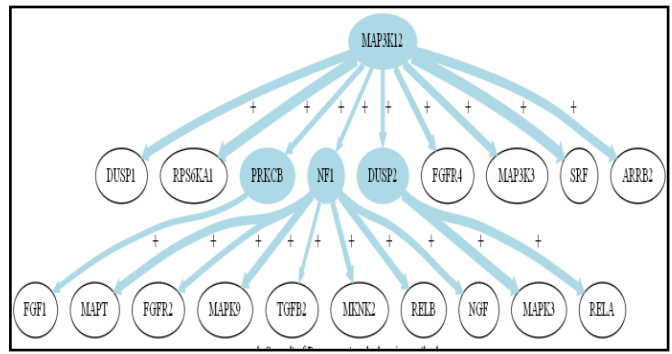

Figure 2: The model resultant from Bayesian network learning method

In Figure 2 the thickness of the edge indicates the amount of causality between genes. The resultant Bayesian network from Appling simple linear regression cannot find how different genes work together to regulate one gene (many-to-one). Furthermore, it is difficult to use another method such as the multiple linear regression to learning Bayesian network to find acyclic graph, because we need the prior knowledge to prevent a cyclic graph which we do not have in our data. Therefore, we will show how Dependency networks are used to find different layers of relationships.

### C. Learning Dependency networks(DNs)

In Learning Dependency networks, we will use correlation coefficient($r$) to select the best five parents for each one of 20 genes that have the highest value of $r$. Then in order to learn the Dependency network based on the selected genes, we used multiple linear regression and constructs the model from the dataset accordingly.

$$Y = \beta_0 + \beta_1 x_{i1} + \beta_2 x_{i2} + \ldots + \beta_n x_{in} + \epsilon_i \, , i = 1, 2, \ldots, n \quad (3)$$

Figure 3 shows the result of applying Dependency network method using multiple linear regression.

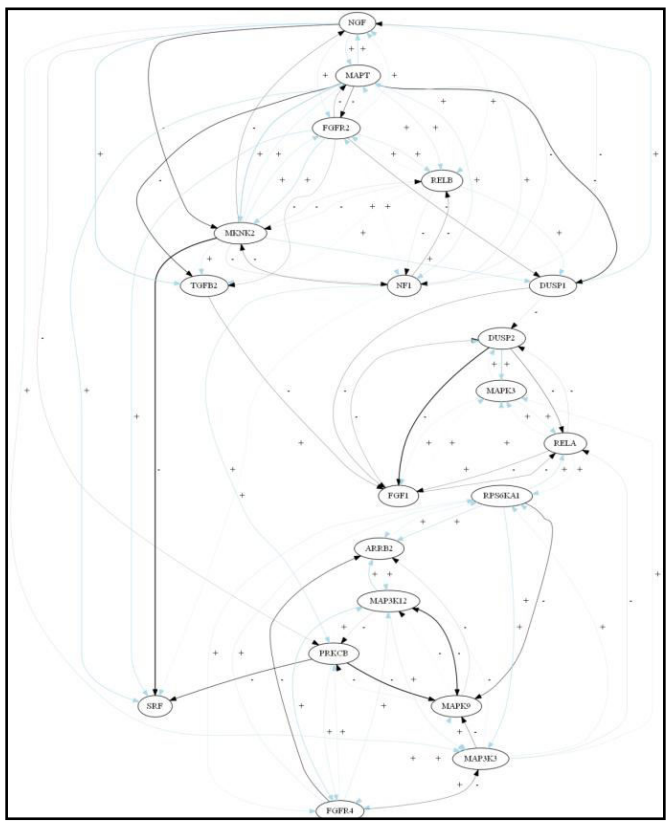

Figure 3: The model resultant from Dependency network learning method

In Figure 3 the level of interaction between genes and its parents represented by the thickness of edge.

### D. Evaluation of our graphical models

To evaluate the models we used leave-one-out cross-validation (LOOCV) in order to estimate the error rate and select the best model in terms of the prediction accuracy. The results show that the overall error for the Co-expression model in Figure 1 is 4.66196. The overall error for the Bayesian model in Figure 2 is 0.925545. The overall error for Dependency model in Figure 3 is 0.0827211**.** Therefore, the Dependency model has the best prediction accuracy comparing with others methods, as we see in Figure 4.

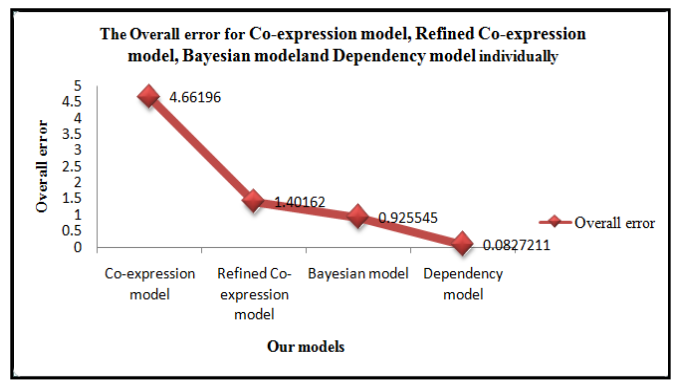

Figure 4: The Overall error for Co-expression network, Refined Co-expression network, Bayesian network and Dependency network individually

### III. CONCLUSIONS AND FUTURE WORK

In the work represented here, we experimented with different Machine Learning of Graphical Models. The results have shown that the Dependency network has the best prediction accuracy comparing with others methods. Therefore, according to the dataset we learned from, the Dependency network has been considered to be the most suitable one to learn the regulations between genes when the breast cancer is developed. In future work, we will experiment with more datasets so that a better idea of the usability of Dependency networks can be seen.

# Using the Data Mining Techniques for Breast Cancer Early Prediction

Samar Al-Qarzaie, Sara Al-Odhaibi, Bedoor Al-Saeed, and Dr. Mohammed Al-Hagery*
*Department of IT, College of Computer, Qassim University; Saudi Arabic

*Abstract*: Data mining (the analysis step of the "knowledge Discovery in Databases" process, or KDD), a field at the intersection of computer science and statistics, is the process that attempts to discover patterns in large data sets. The main objective of this paper is using the data mining techniques and tools for breast cancer early prediction, and to connect the technical field with medical field to serve the community. The research problem is that the specialists are not using the accumulated medical data for prediction purpose. This problem caused the loss of time and effort in hospitals and spending lots of efforts and costs and repeating the same treatments on the same cases. In this paper our focus is to use data mining techniques and tools to extract knowledge from medical datasets which related to breast cancer disease. This dataset collected from the international Repository of Artificial Intelligence. Furthermore, the research used that dataset and applied it in data mining Tools and algorithms. The data mining technique and tool used was the Decision Tree technique and WEKA tool.

Index Terms: Data Mining, KDD processes, Decision Tree, WEKA, Knowledge discovery, and medical datasets.

## I. INTRODUCTION

Currently, data mining techniques are employed to extract knowledge from medical datasets which related to breast cancer disease. Where the intelligent method applied in an essential computer-assisted process to dig through and analyzing an enormous sets of data and then extract specific data patterns in assist of tools to predict behavior along with the future trends allowing the beneficiary to make a proactive, knowledge- driven decision and to answer question that were traditionally too time consuming and expensive to resolve [1].

The overall goal of the data mining process is to extract advanced information from a data set and transform it into an understandable structure for further use [2]. In data mining the strengths and weaknesses of each of the new methods can be demonstrated on data by discussing applications of these techniques to current problems in all domains. Data mining can help in the area of medicine and pharmaceuticals to better determine which patients benefit from a given treatment [1]. The data mining provides information about cancer, including state of the art information on cancer screening, prevention, treatment and supportive care, and summaries of clinical trials [3]. When creating a data mining model, must first specify the mining function then choose an appropriate algorithm to implement the function if one is not provided by default [4]. Several works were done in this field especially in Breast Cancer Diagnosis by other methods such as using k-Nearest Neighbor with different distances and Classification Rules as in[5]. Other methods are used the Classification Based on Associations Rules [6].

The research problem is founding a huge amount of medical datasets are not applied to assists the specialists and professionals in the medical field. The main objective of this paper is to use data mining techniques to extract knowledge from medical datasets.

The rest of the paper is organized as follows. Section 2 presents the technique and tool; section 3 shows the analysis of result followed by conclusions in section 4.

## II. PROPOSED TECHNIQUE AND TOOL

The proposed tools and techniques include WEKA tool and Decision Tree technique.

Data mining tools are software components. The proposed tool that will be applied is WEKA software tool because it is support several standard data mining tasks. WEKA is a collection of machine learning algorithms for solving real-world data mining problems. It is written in Java and runs on almost any platform.

The proposed technique that will be applied in this paper is Decision Tree (C4.5) because it is powerful classification algorithms that are becoming increasingly more popular with the growth of data mining.

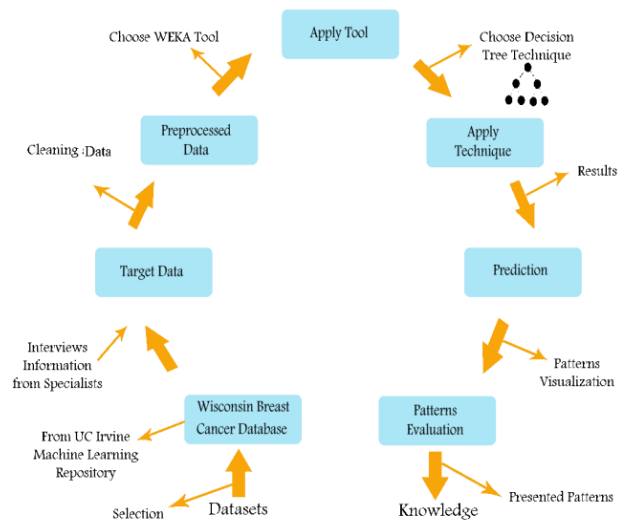

Figure 1: General structure of the research.

Also, the Wisconsin Breast Cancer Database was used. It was collected by Dr. William H. Wolberg, University of Wisconsin Hospitals, and Madison. It has 10 attributes plus the class attribute and 699 instances. The research methodology consists of several steps and sub steps, including the data collection and data analysis and data preprocessing and analysis. The final step is to generate the results, and interprets the final results, as illustrated in figure 1.

### III. RESULTS AND EVALUATION

We separated the dataset that contain 699 instances into two sets are training and testing. The training set contains 500 instances otherwise the testing set contains 199 instances. The results that were concluded are as follows:

The error rate was 6.532% and the accuracy rate was 93.467% for the whole results, kindly see figure 2.

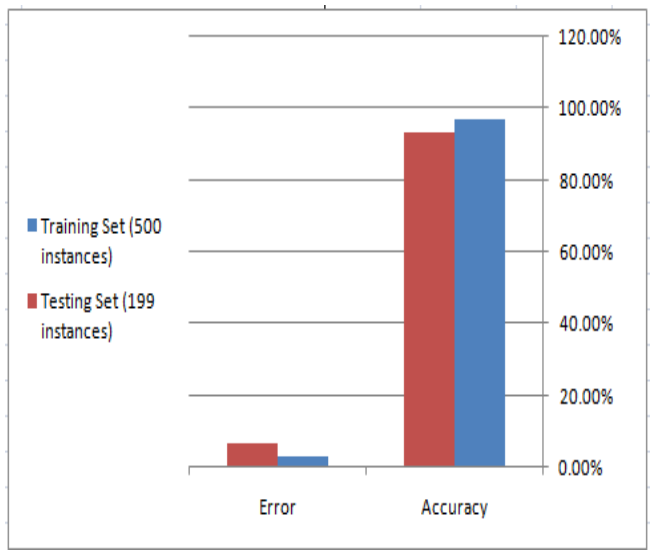

Figure 2: The results obtained from the training and testing sets.

This means that the generated results were more accurate. Thus this is best manner to test predict by WEKA program.

### IV. CONCLUSIONS AND FUTURE WORK

This paper presents an approach for using data mining techniques (decision tree) for the early prediction of breast cancer disease. An important challenge in data mining and machine learning areas is to build precise and computationally efficient classifiers for Medical applications. Our results show that this model is very strong because the proportion of correct in testing set is 93.467% and the proportion of error is 6.532%. Whereas the proportion of correct in training set is 96.8% and the proportion of error is 3.2%. We have achieved in this paper connect the technical field with medical field to serve the community. In addition to that, this work provided practical experiences to reduce time and effort and to assists physician's to predict the disease earlier. In the future, this research work can be intended to find another data sources to use it in the test and prediction. Also, apply other types of algorithms in one time to choose the best way to get the highest accepted values based on updated data sets.

# Correlating new diabetes diagnosis cases for Saudi Population with Google Search Trends

Yasir Javed[1, 2]
[1]Prince Megrin Data Center, College of Computer and Information Sciences, PSU
[2]COINS Research Group, Riyadh, KSA
yjaved@psu.edu.sa

Basit Qureshi[1]
[1]Prince Megrin Data Center, College of Computer and Information Sciences, PSU
Riyadh, KSA
qureshi@cis.psu.edu.sa

*Abstract*—Diabetes is a chronic disease often referred as silent killer that accounts up to 17% of major deaths worldwide. Kingdom of Saudi Arabia is a rapidly developing nation that is suffering from an epidemic of increased diabetes cases accounting to over 40% of total population. Given the improvement in Literacy rates, awareness programs on diseases and widespread use of internet has allowed stakeholders (patients, care taker, health researcher and others) to be more vigilant in searching for relevant information about the disease. Finding exact number of diabetic patients is a time consuming and expensive process that has to be repeated annually. This study tends to answer two questions (1) How search trends can be correlated to total number of diabetic patients (2) what are top search trends in diabetes that can be used in making health related policies and predicting such disease epidemic.

*Index Terms*—Prediction of Diabetes, Google Trends, Correlation of healthcare data.

## I. Introduction

Chronic diseases holds a major share of global disease burden for overall population, it also accounts for major deaths in almost all countries i.e. almost up to 17% according to WHO [1]. Diabetes is one of chronic disease that holds a major share of deaths only in 2013 there were almost 51 Million deaths [2]. Diabetes often referred as silent killer is defined a problem of metabolic disorder. It may be commonly due to glucose intolerance, insulin deficiency and due to its complication patient may often end up dying or some other problems like losing sight, losing legs, losing hands or hooked up on dialysis machine. Diabetes is a major concern as it is becoming major burden on health budgets as it account for almost 1436 USD per patient per year [2]. It is estimated that there will be almost 592 million cases by 2035 that accounts up to 10% population of total world. Urbanization in world is changing the life style, eating habits, work routines mostly in developing countries. KSA (Kingdom of Saudi Arabia) is a rapid developing nation with highly influence of urbanization in life style especially in last two decades. Numbers of diabetic patients are increasing in KSA since last decade due to change in life style and other environmental factors. KSA has a second most number of diabetes cases almost 40% people have diabetes at age over thirty [3]. For diabetic care changes in life style, eating habits and proper education is required; there are plenty of awareness programs for diabetes awareness program like creating Saudi diabetic society, Saudi society for diabetes, KSU, Saudi Diabetic Registry and many other events like conferences, workshops, health camps and brochures. It is evident from internet statistics that usage of internet is continuously increasing that is 0.02 Million users in 2000 to 13 Million users in 2012 approximately 48 times higher [4]. Due to internet usage and literacy program people are tending to use internet for finding information related to disease in terms of disease itself, types of disease, cure about disease and drug related to disease. Most common search engine used in world especially KSA is Google that shares about 58% usages [5]. This study tends to find the common internet searches made for diabetes related terms and its relationship to diabetes in KSA. In KSA, the official language is Arabic while due to huge number of expats above 10 million English is second most common language. This study will focus on search terms both Arabic and English search terms in order to get the real usage scenario with respect to diabetes.

## II. Research Objectives

This Study aims to fill the following aims
1. Finding real time diabetes cases in KSA.
2. Figuring out meaningful search terms used for diabetes in KSA.
3. Finding relationship in between diabetes related search terms and diabetic cases.

## III. Research Methodology

This study spans across two dimensions one is collecting data related to diabetes in KSA for which two sources were adapted one are research papers focusing on search terms like diabetes, diabetes mellitus and type 2 diabetes in KSA. Second source was using International Diabetes Federation (IDF) Diabetes Atlas all six editions to extract real statistics [2] Second aim of this study is to finding diabetic related search trending terms and its usage since 2004 to April 2014, for this Google trends was used as adapted by Carnerio et al [6], Palet et al [7] and chalet et al [8] as this tool provides the searches made during the month for search terms according to countries and region.

For this study data was collected for total number of diabetes patient with number of male and female diabetes patient collected using IDF diabetes atlas. Search terms related to diabetes that was selected according to mostly searched terms focusing disease, its symptoms, its treatments and its medicine. For Arabic and English, separate terms were selected as can be seen in table 1.

| English search terms | Diabetes, Diabetes Mellitus, Diabetes 2, diabetic, diabetes symptoms and Insulin |
|---|---|
| Arabic search term | علاج مرض الـسكري , الـسكر, مرض الـسكري, الأنـسولـين and علاج الـسكر, اعراض مرض الـسكري |

Table1. Indexed search terms used for both English and Arabic.

## IV. ANALYSIS AND DISCUSSION

Details for diabetic patients for year 2003, 2006, 2009, 2011, 2012 and 2013 is presented in Table 1 that shows increase of diabetes over the years. The table focuses on specific age group that is 20 – 79; it shows that number of patients in urban area increasing due to couple of reasons one is development that has converted more areas into urban and other is changes in life style in urban areas.

| Years | Total Population | # of Diabetic Patients | Rural | Urban |
|---|---|---|---|---|
| 2000 | 10,581 | 996.7 | 376.3 | 620.4 |
| 2003 | 10,544 | 992.2 | 70.5 | 921.7 |
| 2006 | 13,730 | 1,854.90 | 140.3 | 1,714.60 |
| 2009 | 15,187 | 2,065.04 | 197.9 | 1,867.14 |
| 2011 | 17,024 | 2,759.56 | 257.6590011 | 2501 |
| 2012 | 17,582 | 3,415 | 265.7317129 | 3148.776995 |
| 2013 | 18,057 | 3,651 | 275.9085777 | 3374.98326 |

Number of people with DM (000's) in the 20-79 age-group

Table 2: Number of Diabetic Patients in Multiple Years with Urban and Rural Distribution using IDF data

Diabetic related search trends in English and Arabic language collected using Google Trends [9] over the span of 2004 to Apr 2014 is shown in figure 1 and figure 2. It is clear from both figures that search trends are increasing in coming years due to various reasons such as increase usage of internet [4], the awareness programs and increase in literacy rate in KSA.

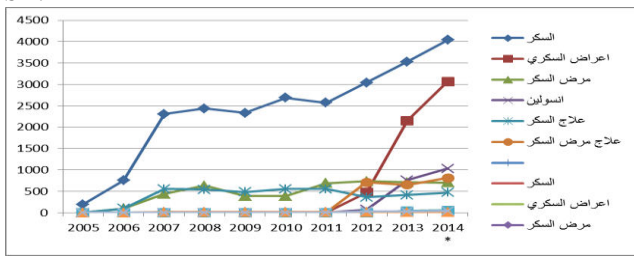

Fig. 1. Showing Search Trends of Diabetic terms in Arabic from 2005 to 2014

It is clear from Figure 1 and Figure 2 that search related to Diabetes related drugs and specific type of diabetes is increasing. Up-going search trends are Insulin, Diabetes 2 and Diabetes Mellitus in English while in Arabic are مرض سكري, علاج الـسكر and الأنـسولـين.

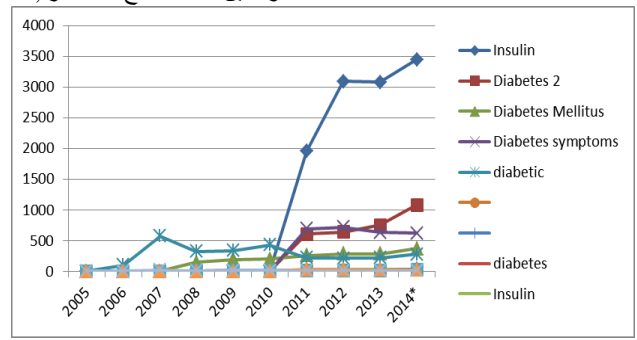

Fig. 2. Showing Search Trends of Diabetic terms in Arabic from 2005 to 2014

In order to correlate the search trends and number of diabetic patients we have took the sum of all search terms in order to include all sort of searches, we took the percentage of diabetic patients according to population

$$\Im = \left(\frac{\Sigma}{\xi} * 100\right)$$

Where $\Im$ is Average Number of Searches and $\Sigma$ is total Sum of All searches in all months and $\xi$ is Total number of Searches in All Years

$$\partial S = (\xi_i / P_i) * \kappa$$

Where $P_i$ is total population $P$ in year $i$, $\kappa$ is constant to be multiplied to make a real mapping between terms and patients and $\partial$ is correlation of trends with Population.

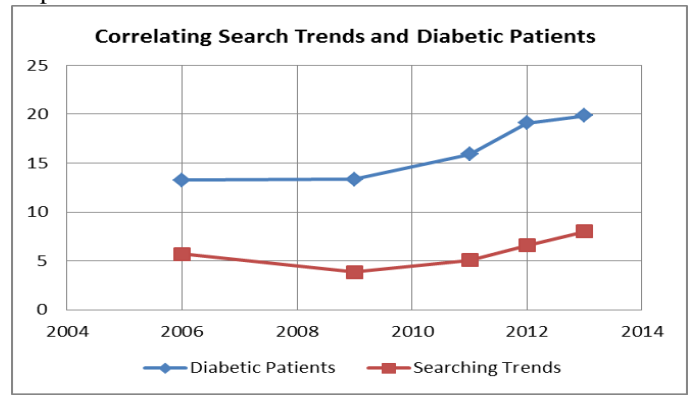

Fig. 3. Correlating search terms with Diagnosed diabetes patients in KSA

Both can fit using a Bi-Nominal equation

$$y = cX^2 - dX + K$$

Where c, d and K are constants and X can be search trend or Number of patients. The Correlation value between both is 0.84291 that shows a high correlation between both sets of data.

V. Conclusion

This Paper focuses on determining the correlation between the numbers of diabetes patients in Saudi Arabia against number of searches on diabetes related search terms. It was observed that both sets of data are 85% correlated to each other, thus providing conclusive proof that internet usage be used to estimate number of future diabetes cases in the county. We also observed that keywords such as Insulin, Diabetes 2 and Diabetes Mellitus in English while in Arabic مرض سكري and الأنسولين ,علاج ال سكر ,علاج are trends that could useful in determining the impact of various awareness programs run by the government that are currently in progress. In future we plan to perform a detailed search term analysis to have a comprehensive understanding about search patterns and their impact on predicting epidemic and chronic diseases.